\documentclass{article}

\usepackage{PRIMEarxiv}
\usepackage{graphicx}
\usepackage{url}
\usepackage[dvipsnames]{xcolor}
\usepackage{tikz}
\usepackage{hyperref}
\usepackage[acronym,nomain]{glossaries}
\usepackage{float}
\usepackage{siunitx}
\usetikzlibrary{arrows.meta,fit,positioning}

\makenoidxglossaries

\newacronym{aps}{APS}{Advanced Photon Source}
\newacronym{alcf}{ALCF}{Argonne Leadership Computing Facility}
\newacronym{epics}{EPICS}{Experimental Physics and Industrial Control System}
\newacronym{hedm}{HEDM}{high-energy diffraction microscopy}
\newacronym{mcp}{MCP}{Model Context Protocol}
\newacronym{iri}{IRI}{Integrated Research Infrastructure}
\newacronym{llm}{LLM}{large language model}
\newacronym{zmq}{ZMQ}{ZeroMQ}
\newacronym{sam3}{SAM3}{Segment Anything Model 3}
\newacronym{rag}{RAG}{retrieval-augmented generation}
\newacronym{eaa}{EAA}{Experiment Automation Agents}
\newacronym{aida}{AIDA}{AI Data Assistant}
\newacronym{pear}{PEAR}{Ptychographic Experiment and Analysis Robot}
\newacronym{apexa}{APEXA}{Advanced Photon EXperiment Assistant}
\newacronym{xas}{XAS}{X-ray Absorption Spectroscopy}
\newacronym{bpilot}{B-PILOT}{Bluesky-Plan Interface for Launch, Operation and Tracking}
\newacronym{bait-mcp}{BAIT MCP}{Bluesky Artificial Intelligence Tools Model Context Protocol}
\newacronym{bits}{BITS}{Bluesky Instrument Template Structure}
\newacronym{bely}{BELY}{Best Electronic Logbook Yet}
\newacronym{moat}{MOAT}{Multi-Office Accelerator Team}
\newacronym{bpm}{BPM}{beam position monitor}
\newacronym{ioc}{IOC}{input/output controller}
\newacronym{fofb}{FOFB}{fast orbit feedback}
\newacronym{pv}{PV}{process variable}
\newacronym{aiops}{AIOps}{artificial intelligence for information technology operations}

\title{Strategies for Deploying AI Agents in Production at Scientific User Facilities}

\author{%
\parbox{0.95\textwidth}{\centering
Ming Du$^{1,*}$, Xiangyu Yin$^1$, Michael Prince$^1$, Yi Jiang$^1$, Rajat Sainju$^1$, Tekin Bicer$^1$, Yanqi Luo$^1$, Eric Codrea$^1$, Peco Myint$^1$, Nina Andrejevic$^1$, Juanjuan Huang$^1$, Trupti Mohanty$^1$, Pawan Tripathi$^1$, Dishant Beniwal$^1$, Hemant Sharma$^1$, Doga Gursoy$^1$, Aileen Luo$^1$, Tao Zhou$^1$, Chenran Xu$^1$, Jan Ilavsky$^1$, Matthew T. Dearing$^2$, Ryan Chard$^3$, Hoon Seo$^1$, Dariusz Jarosz$^1$,
Elaine Chandler$^1$, Thomas Fors$^1$, Hairong Shang$^1$, Pete Jemian$^1$, Xuli Wu$^1$, Madeline Miller$^1$, Ryan Aydelott$^1$, Nicholas Schwarz$^1$, Francesco de Carlo$^1$, Byeongdu Lee$^1$, Yine Sun$^1$, Moises Smart$^1$, Philippe Piot$^1$, Joe Sullivan$^1$, Mathew J. Cherukara$^{1,\dagger}$, Alec Sandy$^1$, Stefan Vogt$^1$, Jonathan Lang$^1$, Laurent C. Chapon$^1$%
} \\[4pt]
  $^1$ Advanced Photon Source, Argonne National Laboratory, Lemont, IL, USA \\
  $^2$ Business and Information Systems, Argonne National Laboratory, Lemont, IL, USA \\
  $^3$ Data Science and Learning, Argonne National Laboratory, Lemont, IL, USA \\
  $^*$ \texttt{mingdu@anl.gov}; $^\dagger$ \texttt{mcherukara@anl.gov}
}

\begin{document}

\textbf{GOVERNMENT LICENSE}

The submitted manuscript has been created by UChicago Argonne, LLC, Operator of Argonne
National Laboratory (``Argonne''). Argonne, a U.S. Department of Energy Office of Science laboratory, is operated under Contract No. DE-AC02-06CH11357. The U.S. Government retains for
itself, and others acting on its behalf, a paid-up nonexclusive, irrevocable worldwide license in
said article to reproduce, prepare derivative works, distribute copies to the public, and perform
publicly and display publicly, by or on behalf of the Government. The Department of Energy will
provide public access to these results of federally sponsored research in accordance with the DOE
Public Access Plan. http://energy.gov/downloads/doe-public-access-plan.

\maketitle

\begin{abstract}
Agentic artificial intelligence (AI) is moving beyond research demonstrations toward production use at scientific user facilities, including light sources, neutron sources, nanoscience centers, and autonomous laboratories. Its scientific value extends beyond increasing throughput. Agents can perform repeatable tasks in calibration, measurement execution, and quality control, as well as initial analyses that turn data into reviewable evidence, allowing scientists to focus on hypotheses, unexpected observations, and interpretation. Drawing on deployments of \gls{llm}-driven agents at the \gls{aps}, this perspective distills practical strategies with an emphasis on elements that can be reused across instruments and facilities. We discuss agent harnesses for beamline control, facility knowledge retrieval, and data analysis while keeping the underlying design principles independent of any specific implementation. These principles cover inference endpoints, tool-server architectures, non-text data, computationally intensive services, reusable skills, and governed learning throughout an instrument's lifecycle. We also consider how network and Linux operations, governed shared memory, and deterministic orchestration can extend these patterns across facility services. Because \glsentryshort{llm} capabilities continue to evolve, these recommendations represent a snapshot of the technology as of the date on the cover.
\end{abstract}

\newpage
\printnoidxglossary[type=\acronymtype]
\newpage

\section{Introduction}

Three converging trends have made agentic AI a credible production technology for scientific user facilities. First, recent \glspl{llm} provide reliable tool use, coding-based problem solving, multimodal reasoning, and the ability to follow task-specific instructions. Second, modern instrument-control stacks expose well-defined programmable interfaces, such as Bluesky/Ophyd connected to the \gls{epics}, through which agents can interact with instruments. Third, the growing adoption of the \gls{mcp} provides a common interface for different harnesses to discover and invoke tools~\cite{Anthropic2024MCP}. Consequently, interactive natural-language control, autonomous execution of bounded procedures, and on-the-fly interpretation of multimodal data are becoming deployable capabilities rather than isolated prototypes. The remaining engineering question is how to deploy these capabilities effectively.

The motivation, however, is not simply to perform more measurements per hour. As instruments and experiments become more capable, scientists devote increasing attention to routine setup, calibration, monitoring, data handling, and recovery from familiar failure modes. An assistant that performs these bounded technical tasks and connects measurements to documentation and validated analysis can shorten the path from data to reviewable evidence, allowing scientists to focus on experimental intent and scientific judgment. When corrections, decisions, and outcomes are retained with their provenance, the system can also accumulate knowledge throughout the instrument operational lifecycle: scientists teach procedures and exceptions to the system, which can later return explanations and operational knowledge to new users. Retrieval-augmented facility assistants, teachable instrument agents, and vision-language experiment automation provide early examples of these complementary roles~\cite{Prince2024CALMS,Vriza2026Teachable,Du2026EAA}.

Several systems have established what agentic AI can provide at scientific facilities. The Context-Aware Language Model for Science (CALMS)~\cite{Prince2024CALMS} demonstrated retrieval- and tool-augmented \glspl{llm} as facility assistants by combining semantic search of documentation with direct instrument-control calls. VISION~\cite{Mathur2025VISION} introduced a modular ``cog'' architecture and reported a voice-controlled experiment at the Complex Materials Scattering beamline of the National Synchrotron Light Source II. Vriza et al.~\cite{Vriza2026Teachable} demonstrated a multimodal, teachable multi-agent system for operating the Hard X-ray Nanoprobe and a robotic synthesis platform, with human feedback retained as retrievable memory. Osprey~\cite{Hellert2026Osprey} packages agentic AI for production deployment in safety-critical control systems. \Gls{pear}~\cite{Jiang2025PEAR} converts natural-language ptychography requests into validated reconstruction scripts, while \gls{eaa}~\cite{Du2026EAA} supports workflows ranging from interactive sessions to deterministic logic with embedded \glsentryshort{llm} queries. These projects address different aspects of agentic operation; here, we consider how to translate their demonstrations into routine practice across the \glsentryshort{aps} and other facilities.

We distill this strategy from deployments that span instrument control at the 26-ID hard X-ray nanoprobe~\cite{Vriza2026Teachable} and the 2-ID-D microprobe endstation~\cite{Du2026EAA}, retrieval from electronic logbooks and technical documents, and analysis of \gls{xas}, \gls{hedm}, and ptychographic data~\cite{Jiang2025PEAR}. Appendix~A collects the named implementations and examples of their interfaces, whereas the main text is organized around the functions that these systems provide. We first present the design principles and then describe recommended practices for inference endpoints, tool servers, data handling, computation, and skill decomposition. We extend these practices to network and Linux operations, shared knowledge and memory, and orchestration across facility services before presenting the worked nanoprobe demonstration. These recommendations are independent of a specific harness. Finally, we identify both immediate facility needs and longer-term opportunities for scientific user facilities.

\subsection{Target architecture}

\begin{figure}[t]
    \centering
    \includegraphics[width=\textwidth]{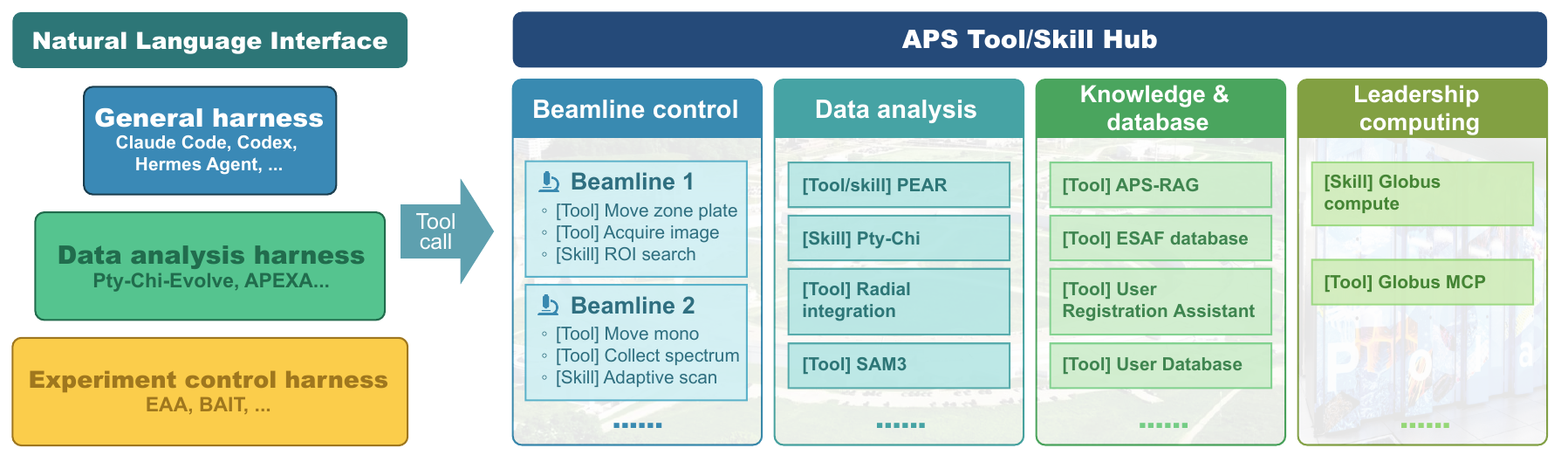}
    \caption{Target capability architecture for agentic AI at the \glsentryshort{aps}. The central \glsentryshort{mcp}/Skills Hub makes instrument control, scientific analysis, knowledge retrieval, and leadership computing available to specialized applications and general agent harnesses. The diagram presents the capability layer; safety, identity and authorization, observability, provenance, evaluation, and lifecycle learning are cross-cutting requirements rather than separate spokes.}
    \label{fig:hub}
\end{figure}

Figure~\ref{fig:hub} summarizes the high-level strategy. A facility-wide \gls{mcp}/Skills Hub, shown here for the \glsentryshort{aps}, organizes four categories of capability:
\begin{enumerate}
\item Knowledge retrieval from machine logbooks, publications, technical reports, experiment records, and other facility sources
\item Instrument and accelerator control across beamlines
\item Online and offline scientific analysis
\item Access to leadership-class computing through services such as the \gls{iri}, Globus Compute, and \glsentryshort{mcp} tools
\end{enumerate}

The hub connects any agent harness that follows the \glsentryshort{mcp} contract to instruments and services that provide or use these capabilities. Computationally intensive workloads outside the instrument environment are attached as external services. Current examples include \gls{sam3} inference and Pty-Chi ptychographic reconstruction on the Polaris leadership-class system at the \gls{alcf}, as well as model training in a commercial cloud. Figure~\ref{fig:hub} therefore presents the capabilities of the architecture rather than a complete operational stack. Identity and authorization, physical safety, observability, provenance, evaluation, and lifecycle learning remain requirements across every spoke. Sections~\ref{sec:design-principles}--\ref{sec:tools-skills} explain how the design principles determine these components, Section~\ref{sec:demonstration} shows how one instrument-control spoke is implemented at the 26-ID nanoprobe, and Section~\ref{sec:outlook} outlines the capabilities that remain to be developed.

The four capability categories also depend on shared network, Linux, storage, and service infrastructure. Sections~\ref{sec:aiops-facility-infrastructure}--\ref{sec:unified-control-plane} extend the architecture to operational support for these dependencies, a governed knowledge and memory service, and orchestration over deterministic execution interfaces. These are cross-cutting extensions rather than additional instrument-specific spokes. A unified control plane should provide consistent discovery, resource identity, and task and result contracts while preserving the authority of each underlying system. A capability catalog is therefore one component of this design, not a substitute for its runtime controls.

\section{Design principles}
\label{sec:design-principles}

\begin{itemize}
    \item Choose an \gls{llm} inference service that is accessible from the computer on which the agentic application is deployed. To reduce operational complexity, we recommend deploying the application on a computer with direct access to the instrument control interfaces. These computers typically reside on an isolated private network. Inference services accessible from devices on this network may include institutionally provided models, \textit{e.g.}, Argo at Argonne, and/or self-hosted local \glspl{llm}.
    \item Decouple the agent harness from low-level control routines and/or data processing packages. Building these tools directly into the harness forces them to share dependencies, making the environment difficult to resolve and maintain. This is particularly problematic with instruments having specialized requirements, such as legacy x86 libraries or older Python interpreters. Coupling tools to a specific harness also limits their reuse by other harnesses.
    \item Ensure interoperability between tools and control libraries. For example, some control libraries manage their own asynchronous event loops or thread pools. Tool integrations should be designed so that thread-safety or asynchronous-execution issues do not crash the program. In addition, parallel tool calls may have unpredictable consequences when they operate real instruments. The best engineering practice is to implement guards that mark instruments or facilities as unavailable and reject further commands issued to busy objects. For example, the accelerator control software has some ``run-control'' process variables (PVs) that indicate hardware being modified by another process, and prevents the launch of new automated processes. However, guards on the agentic harness level should still be implemented to avoid asynchronous or multi-threaded executions of vulnerable equipments.

    \item Promote tool reusability. Tools should be reusable across agent harnesses whenever possible. This goal further motivates decoupling tools from harnesses and using a common standard, such as the \gls{mcp}~\cite{Anthropic2024MCP}, for harness-tool communication.
    \item Unify access without conflating authority. Use reviewed, deterministic adapters to connect disparate systems through explicit request and result contracts. The orchestrator should retain task and dependency state, execution services should enforce local permissions and report operational outcomes, and shared memory should retain scoped evidence and reusable knowledge. An agent's plan, remembered approval, or completion statement must not replace the corresponding execution or authorization record.
    \item Ensure the physical safety of tool execution. Tools must include guardrails that prevent erroneous agent calls from damaging instruments or creating safety hazards. Do not rely on the agent to enforce physical safety boundaries. Unless these boundaries are enforced at the firmware or control-library level, tool developers should implement explicit validation and operational limits within the tool package. Agent actuation requires exclusive control authority and remains subject to immediate operator override. The agent is further restricted by capability gating and command-rate limits, so only approved classes of commands may be issued. 
    \item Use high-level APIs for instrument control tools. Abstract low-level routines when designing instrument control APIs. Keep these routines within tool or software implementations instead of exposing them directly to the agent. This reduces the unpredictability and performance degradation associated with providing numerous low-level APIs, and the same principle applies to both tools and skills. For example, if the agent only needs to acquire images at specified locations, expose an \texttt{acquire\_image} tool that accepts locations, scan steps, and other necessary parameters. Do not require the agent to move motors, control shutters, and collect detector data through a sequence of separate calls. Generic device-level access remains useful as a deliberate escape hatch for diagnosis and setup, but it should be presented as such: where an instrument already defines high-level plans, as a \gls{bits} instrument does. Plans are the intended agent-facing abstraction and per-signal reads and writes are the exception.
    \item Handle non-text data explicitly. Some tools yield images or other high-dimensional data, such as microprobe datasets with X-ray fluorescence or X-ray diffraction information. When the inference interface expects textual tool results, as many OpenAI-compatible chat-completion interfaces do, a viable path to present tool-yielded images to the \gls{llm} should be implemented.
    \item Make agent actions observable and reviewable. Every tool call, its arguments, and its return value should be logged in a structured form that can be later queried. For actions that may affect the data on the host computer systems, or those that change instrument states or otherwise cannot be undone, the harness should support a human-in-the-loop confirmation mode that requires the operator to approve the call before it executes. Observability and confirmability make agentic operation auditable and reproducible.
    \item Treat learning as a governed lifecycle rather than as uncontrolled self-modification. Before an experiment, an agent may retrieve previous procedures and outcomes; during operation, it may adapt within approved bounds; after the experiment, it may propose memories or skill updates based on the decisions, corrections, failures, and outcomes that were recorded. Knowledge should flow in both directions: scientists teach procedures and exceptions, while agents return explanations, evidence, and reusable lessons. This process extends demonstrated approaches for retaining and reusing human feedback~\cite{Vriza2026Teachable}. However, proposed updates should retain their provenance and become authoritative only after evaluation, expert review, versioning, and approval, with a defined path for rollback.
\end{itemize}

\paragraph{From principles to architecture.}
These principles and the deployment requirements developed below map directly onto the architecture in Fig.~\ref{fig:hub}. The need for reachable and resilient endpoints leads to configurable inference services, while decoupling and reusability motivate the hub and its independently deployed tool servers. Concurrency and safety requirements lead to synchronous control workers, interlocks, and approval boundaries; high-level interfaces and explicit handling of non-text data lead to stable tool and data contracts. Finally, computational requirements motivate external analysis and inference services, while observability and governed learning require cross-cutting services for auditing, provenance, knowledge, and skill registries. The deployment practices below implement this mapping one component at a time.

\section{Recommended deployment practices}

\subsection{General infrastructure design guideline for scientific agents}

\subsubsection{Choosing an agent harness}
\label{sec:harness-choice}

The design principles in Section~\ref{sec:design-principles} are deliberately harness-agnostic. They apply whether the harness is \gls{eaa}~\cite{Du2026EAA}, Claude Code, OpenCode, Hermes Agent, VISION~\cite{Mathur2025VISION}, or a custom Python application.

We recommend the following:
\begin{itemize}
    \item If images are not part of the workflow, use a standard agentic harness such as Hermes Agent, Claude Code, Codex, \gls{apexa}, or OpenCode.
    \item Use \gls{eaa} as a starting harness for APS beamline control deployments that require image comprehension and display, as discussed below.
    \item Use \gls{pear} if the focus is purely on ptychography analysis.
    \item Use \gls{apexa} if the focus is purely on \gls{hedm} analysis.
    \item Use Dr.~\glsentryshort{xas} for \glsentrylong{xas} analysis.
    \item If more specialized functionality or interface features are needed, first consider whether an extensible, open-source harness can provide the required core capabilities; otherwise, build a custom harness tailored to the beamline's instrument-control and analysis requirements.
\end{itemize}

For beamline control, try \gls{eaa} first because it already implements the features that the rest of this paper assumes: vision-language reasoning over instrument-generated imagery, two-way \gls{mcp} compatibility (it can consume tools from external \glsentryshort{mcp} servers and expose its own tools as an \glsentryshort{mcp} service), hybrid autonomy that spans fully agent-driven workflows to deterministic logic with embedded \gls{llm} queries, and optional persistent memory across sessions. Beamlines that adopt \glsentryshort{eaa} inherit the system layout described in Section~\ref{sec:demo-architecture} with little additional integration work. They can then extend \glsentryshort{eaa} as needed or use it as a reference for a custom implementation. Persistent session memory provides continuity between interactions, but it does not by itself constitute lifecycle learning. Before information is promoted into shared facility knowledge, it should pass through the provenance, evaluation, and review process defined above.

We treat \gls{eaa} as a starting step rather than a requirement. Claude Code is well-suited to rapid prototyping and workflows in which agent-authored code is the deliverable. A purpose-built harness may be warranted for instruments with unusual safety constraints or hard real-time requirements. As long as the chosen harness honors the principles in Section~\ref{sec:design-principles}, the practices that follow in Sections~\ref{sec:llm-endpoint} through \ref{sec:tools-skills} apply unchanged.

\paragraph{Considering an extensible, open-source harness.}
Before implementing a new harness, teams should consider whether a community-developed harness can provide a maintained core that they extend for local requirements. Pi is one concrete option: its MIT-licensed repository separates the model API, agent runtime, coding-agent interface, terminal UI, and WebUI components into reusable packages. Its agent and coding-agent layers already support persistent session trees, image content in user and tool-result messages, and configurable serial or parallel tool execution~\cite{PiToolkit2026,PiRuntime2026}. These shared mechanisms do not by themselves guarantee reliable beamline operation, which still requires local testing and the safeguards described in this paper. However, they can reduce the amount of facility-specific infrastructure that must be implemented and maintained. Because Pi's development, examples, and issue tracker are public, while extensions can be distributed as npm or Git packages, local additions can also be reviewed and reused outside the original deployment~\cite{PiExtensions2026}.

Pi provides two documented integration paths that permit such additions without maintaining a fork. A Node.js application can embed an agent session through the Pi software development kit, whereas a language-independent application can run Pi in remote procedure call (RPC) mode and exchange newline-delimited JSON commands and events with the process. The RPC protocol also carries extension-generated dialog requests and client responses, allowing a custom WebUI backend to present confirmations or input dialogs and to relay user responses while Pi retains the agent loop and session state~\cite{PiExtensions2026}. Multi-agent behavior can likewise be added as an extension rather than a core modification. For example, Pi's official subagent extension launches separate Pi processes and supports single, parallel, and chained execution, with the output of one agent passed to the next; a generator-to-reviewer sequence can therefore be assembled from the documented chain mechanism~\cite{PiSubagent2026}.

This extensibility introduces practical limitations. Pi extensions are TypeScript modules, which creates an additional implementation and maintenance burden for teams whose software stack is primarily Python. Extensions and installable packages also execute with the user's system permissions, so third-party code must be reviewed and constrained before it is connected to instrument-control tools~\cite{PiExtensions2026}. Moreover, Pi deliberately leaves Model Context Protocol integration, subagents, and background shell execution outside its core~\cite{PiToolkit2026}. The extension example demonstrates that subagents are possible, but conditional or cyclic multi-agent workflows require custom orchestration; in contrast, LangGraph directly represents shared state, nodes, branches, loops, and parallel execution, and documents evaluator--optimizer and orchestrator--worker patterns~\cite{LangGraphWorkflows2026}. Similarly, Pi does not document a built-in mechanism that releases a running tool with a queryable session identifier. Such non-blocking execution would need to be supplied by an external job service or by a custom submit-and-poll tool contract. Teams should weigh these integration costs against the benefit of reusing a shared agent runtime before selecting Pi or developing a purpose-built harness.

\subsubsection{\glsentryshort{llm} inference endpoint}
\label{sec:llm-endpoint}

Two practical issues are worth flagging at deployment time. First, the harness can assume an API format (\textit{e.g.}, the OpenAI Chat Completions API), but should be provider-agnostic and, if possible, model-agnostic. Endpoint URLs, model names, and authentication keys should therefore be configurable. Hard-coding configuration values, especially API keys, should be avoided even when the harness runs under a service account. 

Second, the agent harness should have retry mechanisms that reattempts queries when the initial query fails due to rate limit or temporary server hangs. A primary/fallback mechanism, where the harness attempts the preferred endpoint (for example, Argo) and transparently falls back to an OpenAI-compatible local model when the primary endpoint is unreachable, can be considered, but the switch to the fallback path should be made transparent to the user, or require human approval.

\subsubsection{Handling tool-yielded data}
\label{sec:handling-tool-yielded-data}

When the \gls{llm} provider expects textual tool messages, as in many OpenAI-compatible chat-completion interfaces, the harness must process images and other non-serializable data separately. Even with APIs that support richer multimodal content, harness-side design is still required if the agent needs to access raw numerical data (\textit{e.g.}, NumPy arrays). This section introduces recommended strategies for image viewing and numerical data access.

\paragraph{Letting the agent ``see'' tool-generated images with its vision.}

If the goal is to let the agent ``see'' tool-generated images without performing numerical analysis, two approaches are available:
\begin{itemize}
    \item Let the tool render the image as a PNG file, save it to disk, and return its absolute path. The harness loads the image, encodes it in base64, creates a user or system message with the encoded image, and injects the message into the context. This approach requires the tool server and harness to run on the same computer or use shared storage mounted at the same absolute path. The harness and tool must follow a contract that enables the harness to identify the image path. We recommend returning a JSON object with an \texttt{img\_path} field like the following:
    \begin{verbatim}
        {
            "img_path": "/path/to/image.png"
        }
    \end{verbatim}
    The harness should parse the JSON object and handle images when this field is non-empty.
    \item Let the tool encode the image in base64 and transmit the encoded data directly. The harness moves the encoded data out of the tool response and into a follow-up user or system message. This approach does not require the tool server and harness to access the same storage, but it becomes much more important to ensure the harness can handle it properly when using the Completions API -- in that case, the harness should intercept the tool return, extract and remove the encoded data from the tool response message, then synthesize a user message with the encoded image. If the harness blindly forwards whatever the tool returns to the inference endpoint that uses the Completions API, the image will not be understood and the base64 encoding could consume a large number of tokens. Modern APIs like the Responses API can receive encoded images directly contained in tool responses, but fallback mechanisms should still be implemented if the Responses API is not guaranteed.

    \fbox{\begin{minipage}{\linewidth}
    \textbf{Caution}: The harness and tool must follow a contract that enables the harness to identify the encoded image data, intercept it, and send it to the LLM in a conforming format so that the LLM inference provider recognizes the image. Directly sending the encoding to the \gls{llm} as text should be strictly avoided: \glspl{llm} won't ``see'' base64 encoding as images, and the near-random characters in a base64 encoding uses a very large amount of tokens.
    \end{minipage}}
    
    We recommend returning a JSON object such as:
    \begin{verbatim}
        {
            "encoded_data": {
                "type": "image/png",
                "dtype": "uint8",
                "shape": [1024, 1024, 3],
                "data": "<base64-encoded-data>"
            }
        }
    \end{verbatim}
\end{itemize}

For either case, if the data acquisition logic called by the tool always saves images to disk in a non-image format (for example, an HDF5), the tool should extract the image, render it as PNG, and return. Directly returning the path to the non-image data file and expecting the agent to load and render the image is also possible, but not recommended as it requires additional data extraction and image rendering by the agent. 

\paragraph{Letting the agent access numerical data of images and arrays.}

If the goal is to let the agent perform numerical analysis of images or arrays, the data should be saved to disk and the agent should be given the path to the file. The agent can then write analysis code that loads the data. We consider two scenarios here:

\emph{Scenario 1: the data acquisition backend already saves data to a structured format, such as HDF5}

In this case, let the tool return the path to the data file and expect the agent to load and/or decode data from that file when writing analysis code. However, if the data structure of the file is complicated, the agent should know how and where (\textit{e.g.}, the path to the dataset in an HDF5 file) to find and load the data. We recommend embedding the file paths and dataset paths in the returned payload. For example:
    \begin{verbatim}
        {
            "data_file": {
                "name": "detector_frame",
                "type": "hdf5",
                "path": "/path/to/data.h5",
                "hdf5_dataset_path": "/exchange/data"
            }
        }
    \end{verbatim}

Alternatively, after the backend logic saves the data to disk, load the data to memory in the tool, save only the relevant array as an NPY file. This requires an additional processing step in the tool logic but reduces the tool calls and context usage by the agent.

\emph{Scenario 2: the yielded data is small and the backend logic does not save it to disk by default}

If the relevant data is only one or a few arrays and live in memory (e.g., data streamed back from the detector), they may be transmitted through either file storage or direct encoding:
\begin{itemize}
    \item Let the tool server save the array as an NPY file in storage shared with the harness process, then return the path and other metadata. We recommend a JSON object such as:
    \begin{verbatim}
        {
            "data_file": {
                "name": "detector_frame",
                "type": "npy",
                "path": "/path/to/array.npy",
                "shape": [1024, 1024],
                "dtype": "float32"
            }
        }
    \end{verbatim}
    In this case, the harness can pass the returned object directly to the agent.
    \item Let the tool server encode the array in base64 and transmit it through HTTP. The harness intercepts the returned object, recovers the encoded array, saves the array to disk, and sends the resulting path to the agent.
    \fbox{\begin{minipage}{\linewidth}
    \textbf{Caution}: The harness and tool must follow a contract that enables the harness to identify the encoded image data, intercept it, and send it to the LLM in a conforming format so that the LLM inference provider recognizes the image. Directly sending the encoding to the \gls{llm} as text should be strictly avoided: \glspl{llm} won't ``see'' base64 encoding as images, and the near-random characters in a base64 encoding uses a very large amount of tokens.
    \end{minipage}}
    We recommend a JSON object such as:
    \begin{verbatim}
        {
            "encoded_data": {
                "type": "array",
                "dtype": "float32",
                "shape": [1024, 1024],
                "data": "<base64-encoded-data>"
            }
        }
    \end{verbatim}
    
\end{itemize}

\subsubsection{Tool naming}

For most APIs, the layout of tool schemas is flat---that is, the schemas of different tools are not organized by any kind of hierarchy and all schemas are presented on the same level. As a consequence, tool names must be self-contained and should explicitly include the functionality that a tool is associated with. This is important for better clarity and for avoiding clash of tools with the same names. For example, in an \gls{mcp} server used by the APS 2-ID-D beamline that contains a tool for image acquisition, instead of naming the tool simply as
\begin{verbatim}
    acquire_image
\end{verbatim}
name it as
\begin{verbatim}
    aps2idd.acquire_image
\end{verbatim}
to avoid ambiguity.

\subsubsection{Making long tool executions non-blocking}

Tools that perform data acquisition or complicated analyses may take minutes or
even hours to finish. In a single-threaded, synchronous agentic harness, such a
tool blocks the agent loop and prevents the agent from responding to the user or
performing other tasks until the execution finishes. Allowing selected tool
executions to continue in the background can therefore substantially improve
the user experience.

Simply scheduling a tool as an asynchronous routine is often not sufficient: an
OpenAI-compatible Chat Completions API, for example, expects every tool call in an assistant
message to be paired with a corresponding tool response before the conversation
continues. For example, the following context is valid JSON but would likely be
rejected because tool call \texttt{abcdef} is followed by a user message rather
than a tool response:
\begin{verbatim}
[
  {
    "role": "assistant",
    "content": null,
    "tool_calls": [
      {
        "id": "abcdef",
        "type": "function",
        "function": {
          "name": "acquire_image",
          "arguments": "{}"
        }
      }
    ]
  },
  {
    "role": "user",
    "content": "Meanwhile, check the data at..."
  }
]
\end{verbatim}

A harness can provide non-blocking execution without requiring tool
implementations to adopt an additional asynchronous submission-and-polling
contract. One approach is to add an opt-in release timeout and run eligible
calls in daemon worker threads. The graph-owning thread waits for the configured
timeout after starting a worker. If the call finishes within that interval, the
harness returns its result as an ordinary tool response. If the timeout expires,
the harness instead returns an interim tool response indicating that execution
is still underway, while the worker continues running in the background:
\begin{verbatim}
{
  "role": "tool",
  "tool_call_id": "abcdef",
  "content": "{
    \"job_id\":\"12345678\",
    \"status\":\"executing\",
    \"description\":\"The final result will be provided later.\"
  }"
}
\end{verbatim}

This interim message satisfies the call-response pairing required by the Chat
Completions API. It also explicitly tells the model that the submitted work is
still running and that the final result will arrive later. Once this message has
been added to the context, the agent loop can continue, allowing the model to
respond to the user and initiate other work.

When a released execution finishes, the worker normalizes the result, places a
completion record in a run-scoped queue, and signals the main thread.
The main thread drains that queue and appends a harness-defined system
message containing the completed result. This later system message is not a
second protocol-level tool response for the original call; that call was already
paired with the interim tool response. The completion signal can also wake the
agent while it is waiting for user input, allowing the completed result to be
delivered promptly.

Multiple released tool executions may run concurrently. Each execution should
therefore receive a harness-generated job ID, and that ID should appear in both
the interim tool response and the later completion message. The original tool
call ID remains necessary for pairing the interim response with the assistant's
tool call, but a harness-owned job ID is preferable for correlating the later,
out-of-band completion because it is stable, explicitly model-visible, and
independent of provider-specific tool-call ID generation.

The harness should scope completion delivery to the graph run that initiated the work, so that late results are not injected into an unrelated conversation. However, operations with persistent consequences require execution records independent of that graph run. Suppressing a stale notification must not delete the accepted request, operation identity, cancellation state, or eventual outcome. The in-process mechanism above does not by itself provide recovery after process failure; such recovery requires a durable execution service or equivalent persisted state. After interruption or an ambiguous response, the harness or orchestrator should reconcile with that service before resubmitting work, because an unanswered request does not establish that the operation was never accepted. A cancellation request should likewise not be reported as completed cancellation without confirmation. Memory summaries may reference these records but must not replace them; Section~\ref{sec:unified-control-plane} extends this distinction to cross-service workflows.

An alternative is to split a long operation into a submission tool and a status
query tool. The agent first calls the submission tool, which returns immediately
while execution continues on a server, and later calls the status tool to obtain
progress or the final result. However, this approach requires dedicated tool-side design,
forces the agent to make additional tool calls, and raises the question of when
and how often status should be queried. When existing synchronous tools must be
supported without such changes, keeping the release and completion mechanism in
the harness is generally simpler.

\subsubsection{Deploying computation-intensive tools}
\label{sec:compute-intensive}

Computationally intensive tools, such as neural networks or GPU-accelerated data processors, can be deployed on computers with suitable hardware and HTTP connectivity to the computer hosting the harness. 

\paragraph{Edge devices.}

To isolate the compute load from other activities on the beamline computer, consider using an edge computing device, such as an NVIDIA DGX Spark. The tool should still be exposed as an \gls{mcp} server to maximize compatibility with agent harnesses. If the tool returns images, follow the recommendations in Section~\ref{sec:handling-tool-yielded-data}. If the tool does not perform instrument control, the asynchronous-execution and thread-safety requirements can be relaxed, and the \glsentryshort{mcp} frontend and compute backend do not necessarily need to be separated into two processes as described in Section~\ref{sec:tool-servers-for-instrument-control}.

\paragraph{HPC inference endpoints.}

The \glsentryshort{alcf} hosts several foundation models on its Sophia high-performance-computing system~\cite{tanikanti2025first}, including the general-purpose, prompt-based image segmentation model \gls{sam3}. If image segmentation is needed to support beamline experiments, consider using this service instead of deploying a separate inference instance.

As a concrete example, the 26-ID nanoprobe deployment described in Section~\ref{sec:demo-architecture} offloads its segmentation inference to the \gls{sam3} endpoint hosted by the \glsentryshort{alcf}. A thin \glsentryshort{sam3} client on a computer with public network access sends requests to a \glsentryshort{sam3} inference server running on the \glsentryshort{alcf}; results are written to file-exchange storage and picked up by the local tool server. This pattern---light client on the beamline, heavy compute off-site, shared file system as the handoff---generalizes to any compute-bound tool that does not directly drive hardware.

\subsubsection{Structured session state for task-specific workflows}
\label{sec:session-memory}

Agent harnesses commonly maintain continuity across stateless inference calls by supplying previous messages as part of each request. For a specialized experimental workflow, an alternative is to construct the context from a structured record of the information needed by subsequent decisions. This approach is suitable when the relevant experimental state can be represented explicitly, whereas workflows involving open-ended user interactions or heterogeneous tool use may also require conversation history and detailed tool results. General-purpose harnesses can combine these representations according to the task. As discussed in Section~\ref{sec:harness-choice}, retaining session information provides continuity without constituting lifecycle learning.

In the X-ray fluorescence region-of-interest finder deployed at the Bionanoprobe, we retain the experimental intent, validated control values, selected feedback events, and identifiers of processed scans. Each of the six model jobs in the loop---context interpretation, feedback interpretation, cascade-depth decision, result review, voter evaluation, and box refinement---receives a compact rendering of this record, including the current controls and the last ten events, each capped at a few hundred characters. This representation makes the retained experimental state available across the specialized jobs without requiring each job to receive the complete interaction transcript. Its adequacy depends on whether the retained fields capture the information required by the workflow; constraints or observations omitted from the record are unavailable to later decisions unless supplied separately. An example record is shown below:
    \begin{verbatim}
        {
          "intent": "find individual zinc-rich cells, tightly bounded",
          "state": {"channels": ["Zn", "Ca", "K"], "max_rois": 5,
                    "roi_grouping": "individual", "size_band": [49, 197]},
          "events": [
            {"t": "2026-08-15 14:02:11", "kind": "feedback_text",
             "scan": "bnp_fly0001",
             "detail": "the boxes are too large, I want individual cells",
             "patch": {"roi_grouping": "individual"}},
            {"t": "2026-08-15 14:05:40", "kind": "verdict",
             "scan": "bnp_fly0001",
             "detail": "accept 1 2, reject 3 4 5; banked 2 positive, 3 negative"}
          ],
          "processed": ["bnp_fly0001.mda.h5"]
        }
    \end{verbatim}
The harness persists the structured record to support recovery after a restart and reuse with the fallback inference endpoint described in Section~\ref{sec:llm-endpoint}. Control updates are validated before storage, and the file is replaced atomically after each event, with the event recording the post-change controls. The resume path (\texttt{--resume-session}) restores intent, controls, and recorded history, while processed-scan identifiers allow the workflow to skip scans already recorded as completed. These mechanisms preserve the explicitly recorded state, but do not guarantee equivalent model behavior across endpoints or recovery of information excluded from the record. The event log also provides provenance for the governed write-back described in Section~\ref{sec:tools-skills} and a per-run record for the observability service proposed in Section~\ref{sec:outlook}; promotion of session information into shared skills or memory remains subject to the evaluation and review described in Section~\ref{sec:design-principles}.

Context construction also affects inference efficiency. Although bounding the rendered record limits its contribution to input length, repeatedly rewriting an early prompt block can reduce prefix-cache reuse on endpoints that support prompt caching. Stable instructions and tool definitions should therefore precede changing session information where the API permits. The choice between replaying message history and reconstructing context from structured state should account for information requirements, input length, cache reuse, and measured latency or cost.

\subsection{Instrument-control agents}

\subsubsection{Security boundaries and least-privilege execution}
\label{sec:instrument-security}

An instrument-control agent must be treated as a fallible operator rather than as a safety controller. Prompts and skills can describe constraints, but these instructions do not constitute enforceable safety boundaries because the model may interpret or apply them incorrectly. We therefore recommend defense in depth, in which every requested action passes through controls that are independent of the model's behavior. Physical travel ranges, velocity and acceleration limits, forbidden combinations of instrument states, shutter and detector interlocks, exposure bounds, and command-rate limits should be enforced in firmware or the EPICS/Bluesky control layer whenever possible and otherwise in deterministic tool code. A request outside the permitted operating envelope should fail closed before it reaches the hardware, while existing machine-protection systems and immediate operator override remain authoritative.

Deterministic procedures should likewise be represented as executable interfaces rather than as prompt-only instructions. A repeated calibration, acquisition, or recovery workflow should be wrapped as a high-level tool, a validated external API, or a tested script distributed with a skill. The agent then selects the routine and supplies bounded parameters, while program logic determines the low-level sequence of operations. For instrument-control procedures, scripts distributed with skills should call the same access-controlled tool or control interface as the harness; they should not open independent EPICS connections or generate direct EPICS or Bluesky commands that bypass these controls. The available interfaces should expose only the allowlisted tools, plans, devices, and PVs required for the agent's role, with read and write permissions separated where practical. This design preserves the value of a skill as documentation and orchestration without making compliance with its text the only protection against an unsafe action.
An additional safety guardrail could come from the control system layer. A dedicated \gls{epics} gateway configured for agent use can expose a named subset of \glspl{pv} read-only by default and grant write access only to the channels a particular workflow requires. Because the gateway sits between the agent host and the control system, the restriction holds regardless of which harness, tool server, or script is running, and it cannot be lifted by a change to agent code or by a skill that opens its own connection. A dedicated gateway instance also gives the facility a natural place to log and rate-limit agent traffic separately from ordinary client traffic. This makes the gateway the most robust point at which to draw the read/write boundary, with tool-level checks serving as a second layer rather than the only one.

Least-privilege execution must also extend to the host computer and connected services. Agent-generated code and third-party extensions should run under a dedicated unprivileged account and within an environment that exposes only the resources required by the workflow. On Linux, Bubblewrap can construct a mount namespace in which only selected parts of the file system are visible~\cite{Bubblewrap2026}; experiment inputs can be mounted read-only, while writes are restricted to designated scratch and output directories. Because Bubblewrap provides isolation mechanisms rather than a complete security policy, its namespace and mount configuration should itself be treated as security-relevant deployment code. Individual tool servers may instead run in containers, and the entire harness can be containerized when the agent requires general shell or code-execution capabilities. In either case, the deployment should mount only the required files, devices, and control sockets; apply appropriate limits to processes, memory, and compute time; and permit network access only to approved inference, tool, and data services. Credentials should remain in the service or tool that needs them whenever possible, rather than being exposed to the agent. Local services can bind to a loopback or private interface, while remote calls should be authenticated and authorized. These measures limit the consequences of erroneous code execution, but they do not replace access controls and safety interlocks at the instrument.

Resource isolation alone does not prevent two legitimate clients from commanding the same instrument at the same time. Every state-changing access path should therefore participate in a common lock, implemented in the tool or control layer and preferably enforced at the closest point shared by agents, GUIs, notebooks, and automation scripts. Only the caller that holds the lock should be permitted to actuate the protected instrument; competing requests should be rejected or queued until the lock is released. The lock should record its owner and scope, expire safely if the holder fails, and support explicit release and operator revocation. Harness-level guards remain useful for preventing parallel calls within one agent session, but they cannot protect against a second client that bypasses the harness. The single-operation RunEngine used by the Bluesky Queue Server, discussed below, provides this serialization for requests that pass through the Queue Server. However, direct EPICS clients remain outside that boundary unless the lock is also enforced in the common control layer or firmware.

Finally, preventive controls should be paired with confirmation, auditing, and software-governance measures. Consequential or irreversible instrument actions, as well as operations that modify host data, should require human confirmation at the harness. Structured logs should record the authenticated caller, selected tool, validated arguments, result, lock state, and any operator override. Tool and skill deployments should use reviewed, versioned releases, with authenticated promotion and a path for rollback. Together, these records make failures diagnosable and prevent unreviewed extensions, skills, or agent-generated memories from silently changing production behavior.

\subsubsection{Tool servers and integration with control systems}
\label{sec:tool-servers-for-instrument-control}

We recommend decoupling tools from agent harnesses. To reduce dependency conflicts and promote reuse, implement each tool server in its own repository, give it an independent set of dependencies, and run it in a dedicated environment.

It is important to avoid asynchronous-execution and thread-safety issues that can emerge when instrument-control libraries, such as Bluesky and Ophyd, are called directly from asynchronous functions. While common \gls{mcp} libraries such as FastMCP and web frameworks such as FastAPI allow synchronous routines, they may still run an event loop underneath, especially for HTTP, SSE, or streamable HTTP transports. We suggest a clear separation between asynchronous routines and routines that call control libraries; the latter, if implemented in the tool, should be completely synchronous for maximal safety.

\begin{figure}[t]
    \centering
    \includegraphics[width=1\textwidth]{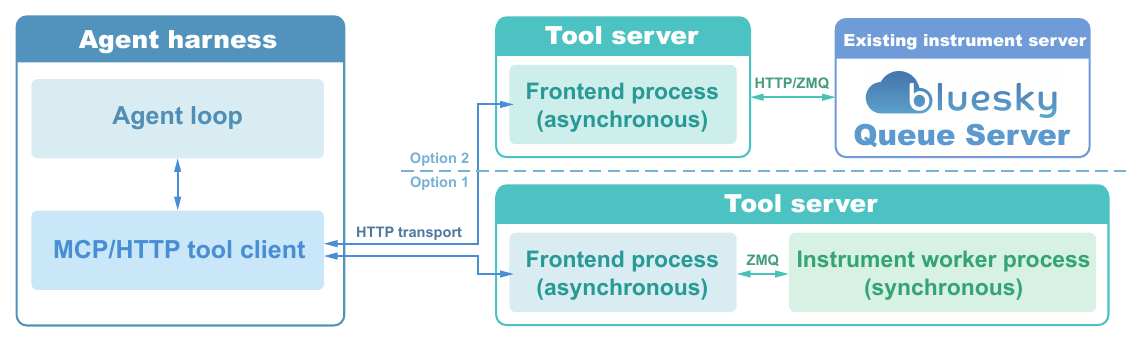}
    \caption{Recommended integration of an agentic harness with an instrument-control tool server designed for asynchronous-execution and thread safety. The key design principle is to separate the asynchronous \glsentryshort{mcp} or HTTP frontend from the routines that perform instrument control. The \glsentryshort{mcp} frontend may handle asynchronous communication with the agent process. Routines that make the actual calls to control libraries (Bluesky, Ophyd, \textit{etc.}) should be placed in a separate worker process that is synchronous, blocking, and communicates with the frontend through \glsentryshort{zmq}. If a decoupled instrument-control server such as Bluesky Queue Server is already available at the beamline, the tool server can communicate directly with that instrument server from its asynchronous routines. \Glsentryshort{bait-mcp} is a deployed instance of this case: it consists solely of the \glsentryshort{mcp} frontend, with the Queue Server serving as the instrument worker.}
    \label{fig:mcp-pattern}
\end{figure}

Fig.~\ref{fig:mcp-pattern} illustrates the possible implementations of a tool designed to support safe inter-operation with instrument-control libraries. In the most basic scenario, where instrument-control routines need to be implemented in the tool itself, the tool server consists of an \gls{mcp} frontend and an instrument worker backend.
\begin{itemize}
    \item The frontend exposes tool functions through the \gls{mcp} interface and may use asynchronous operations.
    \item The backend implements instrument control routines using Bluesky, Ophyd, or other control libraries; all instrument-control calls must remain \emph{blocking} and \emph{synchronous}.
\end{itemize}
The frontend and backend communicate through \gls{zmq}. The backend may run a loop that polls for frontend requests and executes each instrument-control routine synchronously upon receipt. This design is not specific to any control library. The instrument-worker backend may use any suitable library as long as its calls remain blocking and synchronous.

The tool server should depend only on \gls{mcp}/FastMCP, \gls{zmq}, the selected instrument control library, and other necessary packages. It should not share dependency requirements with the agent harness. For more restrictive environments (for example, when instrument control requires Python 2.x and is incompatible with frontend dependencies such as \glsentryshort{mcp}) the frontend and backend may be placed in separate repositories while retaining \glsentryshort{zmq} as the communication layer. Alternatively, the frontend may be implemented as a simple HTTP server without \glsentryshort{mcp}. This approach requires an HTTP client in the harness, wrapped as an agent-callable tool, that queries the server when invoked. It also requires the tool schema to be defined on the client side unless the server and client adopt a contract that allows the client to retrieve schemas automatically. This alternative reduces tool-server portability because not every harness recognizes such contracts.

\subsubsection{Integration with existing servers and GUIs}

The tool-server design pattern can be relaxed if a standalone instrument server is already available at the beamline, so that the tool server may simply post to the instrument server instead of calling the control libraries itself. For example, the Bluesky Queue Server hosts the instrument-control routines and is called via \gls{zmq}. As such, it can be regarded as a replacement for the worker backend in the design introduced above. If the Queue Server is available at the beamline, the \gls{mcp} frontend can communicate directly with the Queue Server via \glsentryshort{zmq}, and the \glsentryshort{mcp} server repository no longer needs its own worker backend. A similar scenario arises when the beamline already has a running server that listens for EPICS commands over HTTP or \glsentryshort{zmq} and executes them on behalf of the caller. The \glsentryshort{mcp} frontend can also request the EPICS server directly instead of maintaining its own backend.

\begin{figure}[t]
    \centering
    \includegraphics[width=1\textwidth]{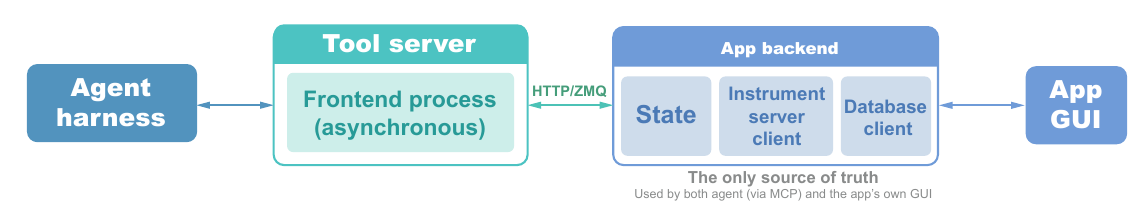}
    \caption{Recommended integration of an agentic harness with an existing application that has a GUI. This design requires the app to have a clean separation into a frontend (GUI) and a backend. The backend implements instrument control and/or data retrieval routines and maintains a state object. The state object is used by both the app's GUI and the \glsentryshort{mcp} server as the only source of truth. User may read or write the state from either the agent (via the \glsentryshort{mcp} server) or the app's GUI, while information stays synchronized between both sides.}
    \label{fig:tool-server-integration}
\end{figure}

If some control capabilities or visualization interfaces have already been realized through an existing application with a GUI, the \gls{mcp} server and the app can cooperate if the app can be cleanly separated into a frontend (GUI) and a backend. The backend implements the instrument control (\textit{e.g.}, communicating with the Queue Server, or running its own control routines) or data retrieval capabilities. It also maintains and exposes a state object that can be read or written by both the app frontend and the \glsentryshort{mcp} server. The state owned by the app backend effectively becomes the only source of truth, so that the \glsentryshort{mcp} server and the app stay synchronized and consistent on instrument status. The Bluesky Queue Server could well serve the role as the app backend if the GUIs at the beamline already listen and post to it.

The Bluesky Queue Server offers an additional benefit of being non-blocking: the native reconciliation of the server gateway and the control engine allows the Queue Server endpoints to be asynchronous, being able to listen to other queries while executing a Bluesky plan. With this, the developer of the tool or harness can implement an asynchronous path that allows the user to pause or stop an ongoing operation, or to check its status from the agent interface.

The closed-loop experimentation setup at 12-ID runs key modules as micro-servers, covering the wet-chemistry platform, robotic sample handling, data acquisition, data reduction, post-processing, and real-time CNN analysis. Each module exposes a minimal set of commands, with fine-grained control accessible via a GUI front end. This architecture isolates the main orchestration loop from single-module failures while simultaneously monitoring all server activity. 

\paragraph{Example: B-PILOT and AutoPILOT at APS beamlines.}
An instantiation of this GUI-integration pattern is under development for the beamlines at Sectors 1, 20, and 3. \gls{bpilot} is a standalone graphical application for building, queuing, and monitoring Bluesky plans: it discovers the plans and devices available for a given beamline through per-beamline configuration profiles, generates parameter-entry forms directly from each plan's signature and docstring, and dispatches runs through the beamline's Bluesky Queue Server, which owns the instrument state and acts as the
single source of truth in the sense of Fig.~\ref{fig:tool-server-integration}. AutoPILOT is an optional agentic layer embedded in \gls{bpilot} as a chat panel. It is a multi-turn \gls{llm} agent that shares \gls{bpilot}'s backend state and, through read-only lookup tools, enumerates the loaded devices and plans, answers questions about them, and drafts plan commands that the user reviews before they are queued or run---keeping a human in the loop for every state-changing action,
per the confirmation principle in Section~\ref{sec:design-principles}. The agent is configured to the Argo endpoint. Importantly, AutoPILOT is a guarded, optional import: \gls{bpilot} runs fully standalone without it, so the interface remains usable when no inference endpoint is available, while the GUI and the agent read and write the same backend state when it is. A planned \gls{mcp} frontend will additionally expose \gls{bpilot}'s backend as a hub-registered tool server, letting external harnesses consume the same instrument-control capabilities. Fig.~\ref{fig:screenshot_bpilot} in Appendix~A shows the \gls{bpilot}
interface with the embedded AutoPILOT panel.
\paragraph{Example: \glsentryshort{bait-mcp} and \glsentryshort{bits} at APS beamlines.} \Gls{bits}~\cite{Codrea2026BITS} is the template structure used to build Bluesky instruments at
APS. A \glsentryshort{bits} instrument repository defines its devices and plans through
declarative configuration, and its \texttt{startup.py} constructs a Guarneri
\texttt{oregistry} that maps device names to live ophyd objects in the RunEngine worker
namespace. The same repository runs in a console session, a notebook, or under a Bluesky
Queue Server without modification. Dozens of APS beamlines now maintain such a
repository. The consequence for agentic deployment is significant: because every
\glsentryshort{bits} instrument presents the same registry structure, the same permissions file,
and the same Queue Server control surface, a \emph{single} tool server can serve all of
them. Standardizing the instrument layer is what turns per-beamline agent integration
into a one-time cost.

\glsentryshort{bait-mcp}~\cite{Codrea2026baitmcp}. is that tool server. It is an \gls{mcp} frontend and a single
\gls{zmq} client of the Queue Server RE Manager, and nothing else: it instantiates no
ophyd devices, opens no EPICS channels, and runs no second device session. This is the simplest form of the pattern in Fig.~\ref{fig:mcp-pattern}, in which the Queue Server \emph{is} the instrument-worker backend and the tool server retains only the asynchronous frontend. Device reads and writes execute in the Queue Server's live session; plans are enumerated, enqueued, and executed through the queue. The server binds to loopback by default and exposes a streamable-HTTP \glsentryshort{mcp} endpoint, so any harness that speaks \glsentryshort{mcp} can consume it. Its tool surface is summarized in

Table~\ref{tab:bait-tools}.

\begin{table}[H]
\centering
\caption{Tools exposed by \glsentryshort{bait-mcp}. Every tool returns a JSON object with an
\texttt{ok} flag, so unmet preconditions (Queue Server unreachable, environment closed,
unknown device) surface as failed tool calls carrying an explanatory \texttt{error} field
rather than as exceptions or hangs.}
\label{tab:bait-tools}
\begin{tabular}{p{0.34\textwidth}p{0.58\textwidth}}
\hline
\textbf{Tool} & \textbf{Purpose} \\
\hline
\texttt{read\_device} & Read a device value; runs in the background, so it is valid
during a plan. \\
\texttt{set\_device} & Set a device value; runs in the foreground and actuates
immediately. \\
\texttt{list\_devices}, \texttt{describe\_device} & Enumerate and describe the devices
visible to the caller's user group. \\
\texttt{list\_plans}, \texttt{describe\_plan} & Enumerate plans and retrieve a plan's
parameter signature. \\
\texttt{queue\_status} & RE Manager state, running item, and queue depth. \\
\texttt{add\_plan}, \texttt{start\_queue}, \texttt{stop\_queue} & Enqueue a plan and
control queue execution. \\
\texttt{run\_plan} & Execute a plan immediately, bypassing the queue. \\
\hline
\end{tabular}
\end{table}

The safety interlock comes from how the Queue Server executes each kind of request. \Glsentryshort{bait-mcp} sends device reads and device writes to the RE Manager
differently. A read is submitted as a background task, which does not compete for
the RunEngine, so an agent can query a motor position or a detector value while a
plan is running without disturbing it. A write is submitted in the foreground,
where the RunEngine executes it in the same single slot it uses to run plans.
Because that slot holds only one operation at a time, a write issued while a plan
is in progress has nowhere to execute and is refused by the Queue Server rather
than racing the plan. The agent is thus prevented from actuating hardware mid-scan
by the control system itself, not by its own restraint. Two limits on that
guarantee should be stated plainly. First, the interlock is scoped to the Queue
Server's own RunEngine, not to the hardware: it can only refuse operations that
pass through the RunEngine, so it cannot see a manual \texttt{caput}, a second
EPICS client, or an operator jogging a motor. \Glsentryshort{bait-mcp} assumes the
Queue Server is the sole controller of the instrument, which is the standard Bluesky Queue Server operating model. Second, \texttt{set\_device} and \texttt{run\_plan} actuate on call. \Glsentryshort{bait-mcp} implements no human-in-the-loop gate by design. Approval is a harness-level policy decision, and duplicating it in every tool server would make it inconsistent across servers and invisible to the operator, who is at the harness. The consuming harness must therefore supply the confirmation mode described in Section~\ref{sec:design-principles}. This division is deliberate, but it is a contract: a harness that consumes \glsentryshort{bait-mcp} without implementing that gate will actuate hardware without review.

\subsubsection{Decomposing beamline procedures into skills}
\label{sec:tools-skills}

Although skills~\cite{Anthropic2025Skills} are becoming increasingly popular, they do not replace tools. A key advantage of running a tool as a server is easier runtime management: the tool server can run on a different computer, with its own dependencies and environment. Achieving the same capability solely with skills is possible, but it requires more agent-driven procedures, longer execution paths, and higher risk. A well-designed tool is also safer and more deterministic because low-level routines are fixed and controlled by program logic rather than provisionally generated by the \gls{llm}. However, because tool schemas are injected into the context, large numbers of tools or excessively long schemas can consume substantial context and degrade performance. To combine the strengths of skills and tools while avoiding their shortcomings, we recommend the following practices:
\begin{itemize}
    \item Provide only high-level abstractions as tools whenever possible. Avoid unnecessary low-level APIs when they can be wrapped together.
    \item Present long, detailed documentation as skills instead of placing it in tool docstrings. Load skill text only when needed. Use skills to provide documentation and instructions that explain how the agent should use the tools.
    \item Skills may include executable scripts or coding instructions, but avoid scripts or APIs that require dependencies beyond those already available in the harness. Dependency-heavy executables should be deployed as tool servers rather than wrapped in skills. If there is a truly unavoidable reason to not use tools for scripts or APIs with extra dependencies, consider including a local, isolated \textit{uv} or \textit{pixi} environment in the skill directory that the agent can activate.
\end{itemize}

The harder question is not whether to use skills, but how to decompose beamline work into them. We use the following framework at APS.

\textbf{Granularity rule.} One skill corresponds to one coherent multi-step workflow that an experienced beamline scientist would describe as a single procedure: ``align the zone plate,'' ``acquire a line scan around feature X,'' ``submit and post-process a ptychography reconstruction.'' Operations that decompose into a single hardware action (move a motor, read a counter, or acquire an image) remain tools, not skills. The test we apply is whether removing the skill would force the agent to re-derive the procedure from raw tool calls every time. If so, the procedure deserves a skill.

\textbf{Composition.} Skills may invoke other skills, but the underlying instrument-control actions should remain \gls{mcp}-exposed tools. A skill is documentation and orchestration; it must not embed instrument-control logic that bypasses the tool server. This avoids the risk of letting the agent freestyle-code instrument-control commands directly with EPICS or Bluesky.

\textbf{Loading on demand.} The harness should expose skill metadata (name, one-line description, applicable instrument) to the \gls{llm}, but load the full skill body only when the skill is invoked. This scales to dozens or hundreds of skills without inflating the prompt, and mirrors the on-demand pattern recommended above for tool schemas.

\textbf{Naming, versioning, audit.} Skills are code: they deserve kebab-case names, semantic versioning, and a review workflow. Skills that drive motors or otherwise change instrument state should pass an additional level of review before deployment.

\textbf{Lifecycle learning and knowledge governance.} The skill registry should support a controlled cycle in which observations are captured, proposed as updates, evaluated, reviewed by scientists, promoted as versioned records, and subsequently monitored or rolled back. Scientists should be able to teach procedures, constraints, exceptions, and diagnostic cues, while agents may propose updates based on logged sessions. For each update, the registry should preserve the source scientist, instrument context, supporting evidence, conditions of applicability, and review status. This process allows accumulated experience to become durable knowledge without permitting unvalidated memories to alter production behavior. Section~\ref{sec:knowledgeops-orchestrated-pipelines} extends this lifecycle to shared evidence, operational episodes, and derived state across agents and runs; Section~\ref{sec:unified-control-plane} separates their reuse from execution authority.

\subsubsection{Worked example: a minimal nanoprobe skill set}

The following example illustrates the combination of tools and skills used at the 26-ID Hard X-ray Nanoprobe:
\begin{itemize}
    \item Tool -- \texttt{acquire\_coarse\_scan}: acquire a coarse scan at the specified location.
    \item Tool -- \texttt{acquire\_fine\_scan}: acquire a fine scan at the specified location.
    \item Tool -- \texttt{segment\_image}: segment an image using \gls{sam3} with a prompt. This tool is a proxy of the ALCF \glsentryshort{sam3} inference endpoint.
    \item Tool -- \texttt{render\_image\_for\_agent}: render a TIFF or NPY file to a PNG image.
    \item Tool -- \texttt{execute\_python\_code}: execute Python code. The caller may specify that the last printed line is an image path, in which case the harness fetches the image and sends it to the \gls{llm} in a follow-up message.
    \item Skill -- \texttt{roi-search-workflow}: documents the problem statement, information needed, procedures, and hints for searching for a user-described feature of interest in the sample.
    \item Skill -- \texttt{zone-plate-focusing}: documents the calibration procedure, sharpness metric, and acceptance criteria, and explains how to use the movement and image-acquisition tools needed to focus the optic.
    
\end{itemize}
The same set of tools and skills compose to support the interactive session in Section~\ref{sec:demo-session}. Together, these tools and skills allow the agent to perform bounded calibration, acquisition, and initial feature localization while the scientist retains control over the scientific objective, interpretation, and consequential decisions.

\subsection{Scientific data-analysis agents}

\subsubsection{Integration with deterministic analysis software}

Scientific data-analysis agents should orchestrate validated analysis software rather than ask the \gls{llm} to perform numerical analysis directly. In this architecture, deterministic tools perform the numerical operations, while the \glsentryshort{llm} plans, explains, and checks the workflow. The agent can therefore provide a conversational orchestration layer over established analysis software while preserving the interpretability, editability, and auditability expected by expert users. This division of responsibilities allows agents to help convert raw measurements into reviewable scientific evidence without using the language model as a substitute for validated numerical methods or expert interpretation.

\subsubsection{Reviewable scientific artifacts and provenance}

The deployment practices above address the backend structure required for reusable and safe agentic systems, but production adoption also depends on the user-facing scientific interface. In many analysis workflows, especially those involving model-dependent interpretation, users will not trust an agent solely because it returns a final answer or reports that a tool call completed successfully. The agent should expose intermediate scientific state in forms that are familiar to domain experts and compatible with existing analysis practice.

Accordingly, analysis tools should return not only machine-readable results, but also structured, reviewable scientific artifacts. These artifacts may include plots, parameter tables, fit reports, project files, provenance records, and exportable data products.

\subsubsection{Progressive autonomy and expert oversight}

In practice, scientific analysis harnesses should support progressive autonomy. A user may first ask the agent to suggest an analysis plan, then approve specific tool calls, inspect intermediate results, modify constraints or fit ranges, and rerun only the affected step. Such interaction patterns make agentic systems more compatible with existing expert workflows and reduce the risk that automation obscures scientific judgment.

\subsubsection{Worked example: \glsentryshort{xas} analysis with Dr.~\glsentryshort{xas}}

\begin{figure}[t]
    \centering
    \includegraphics[width=0.9\textwidth]{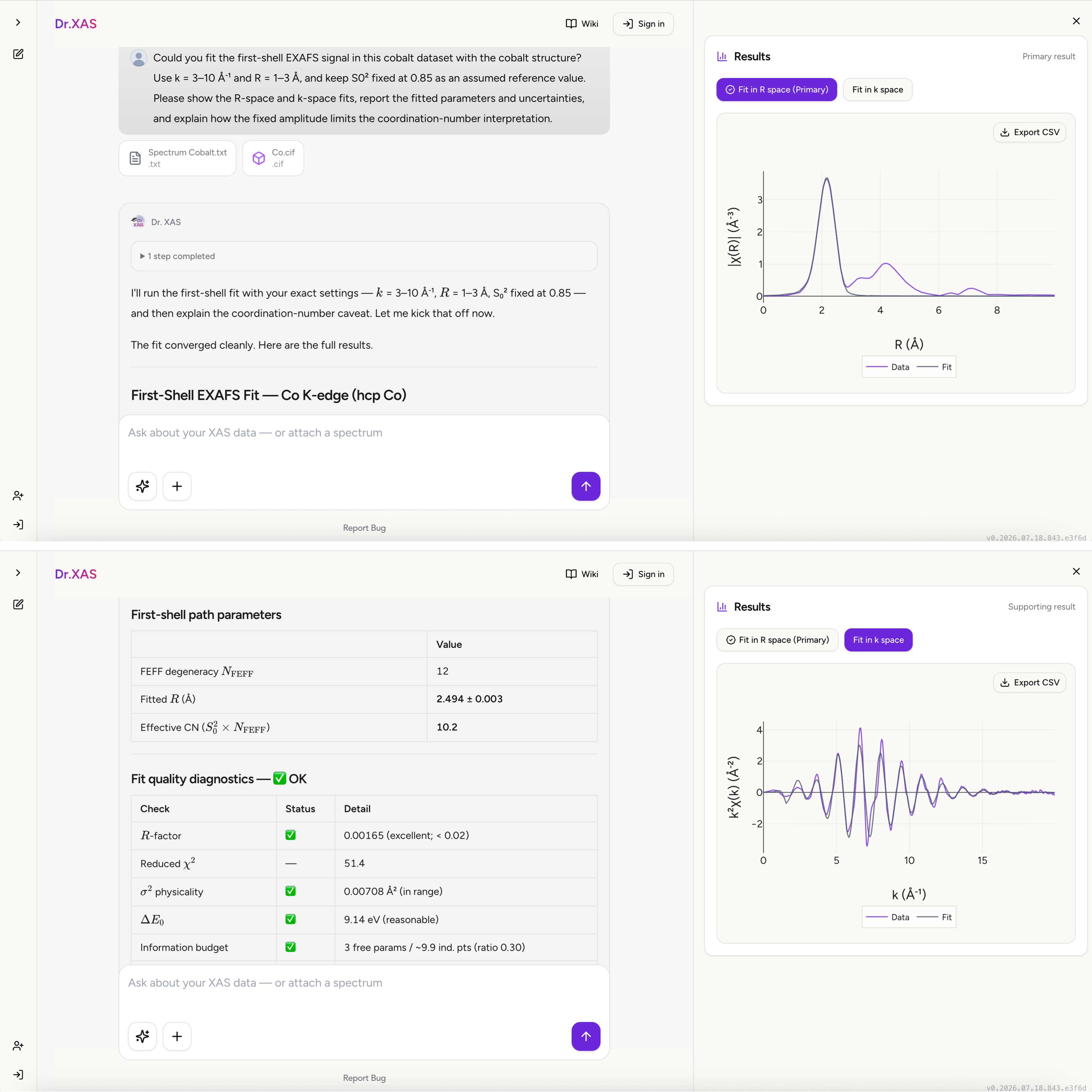}
    \caption{User-facing artifacts in an agentic \glsentryshort{xas} workflow. Dr.~\glsentryshort{xas} exposes conventional EXAFS analysis outputs, including R-space and k-space fits, fitted parameters with uncertainties, fit-quality diagnostics, and interpretation caveats associated with fixed model assumptions such as $S_0^2$. The example illustrates that scientific analysis agents should return reviewable artifacts and assumptions, not only final text answers or completed tool calls.}
    \label{fig:drxas-artifacts}
\end{figure}

\Glsentrylong{xas} provides a useful example. In an agentic \gls{xas} workflow, executable tools may perform deterministic operations such as spectrum loading, normalization, XANES reference comparison, EXAFS signal extraction, FEFF path generation, least-squares fitting, and report generation. However, the user-facing response should also display the assumptions and artifacts needed for expert review: fit ranges, fixed or constrained parameters, fitted parameters with uncertainties, diagnostic metrics, residuals, and plots in the conventional representations used by the community. Fig.~\ref{fig:drxas-artifacts} illustrates this principle for an EXAFS fitting workflow in Dr.~\glsentryshort{xas}, where the agent returns both conventional fit artifacts and the assumptions needed to interpret them.

\subsubsection{\texorpdfstring{Worked example: \glsentryshort{hedm} analysis with \gls{apexa}}{Worked example: HEDM analysis with APEXA}}
\label{sec:apexa-example}
\gls{apexa} orchestrates the MIDAS crystallographic pipeline
(calibration, azimuthal integration, Rietveld/GSAS-II refinement, grain mapping) behind a natural-language interface. The harness follows a multi-agent pattern: a deterministic keyword router dispatches each user query to one of five scoped specialists (Calibration, Analysis, Knowledge, Motor, Visualization), each of which sees only the subset of \glsentryshort{mcp} tools relevant to its remit. Sixty tools are exposed across three \glsentryshort{mcp} servers (core file operations, MIDAS analysis, and EPICS motor control). Because the tool registry is scoped per specialist, the Motor agent cannot invoke destructive file-write or long-running-workflow tools, providing safety-by-construction rather than safety-by-prompting. A screenshot of the \gls{apexa} WebUI is shown in Fig.~\ref{fig:APEXA-GUI}.

\subsection{AIOps for network and Linux operations}
\label{sec:aiops-facility-infrastructure}

\subsubsection{Operational support across service dependencies}

The deployment principles for instrument agents also apply to the network and Linux infrastructure supporting acquisition, analysis, and data movement. We use \gls{aiops} here to describe agent-assisted investigation, change assessment, and recovery of this infrastructure. The object of an investigation should be the affected service and its dependencies, rather than only the host or instrument reporting the symptom. For example, an acquisition failure may require observations from the acquisition process, its host, the network path, and the destination storage service. Agents should return a reviewable account of the affected resources, supporting observations, unresolved alternatives, and the next diagnostic or operational action. The recommendations in this and the following two subsections extend the preceding deployment patterns; they do not constitute a report of a facility-wide autonomous operations deployment.

\subsubsection{Evidence-directed diagnosis}

For recurring investigations, required evidence collection should be enforced by the workflow, while additional hypothesis exploration remains agent-directed. A host-versus-network investigation should resolve the relevant host interface and network attachment, compare observations over compatible intervals, and consult the applicable configuration and change records. Documentation explains how an observation may be interpreted; it does not establish the current state of the system. Similarly, a successful tool invocation does not establish that the required observation was obtained. Workflow state should distinguish usable evidence, a completed observation with no matching finding, unavailable instrumentation, restricted access, and collection failure. A missing observation should remain an explicit limitation rather than be counted as a satisfied diagnostic requirement.

This distinction is particularly important when evidence spans systems with different sampling intervals or retention policies. A recent host event and an older switch snapshot should not be treated as simultaneous observations, and an empty error log should not establish health when its collection path is unavailable. When several hosts exhibit related symptoms, the workflow should examine shared dependencies before proposing separate repairs. The useful exit condition is a supported finding or the next observation needed to distinguish the remaining explanations, not merely completion of a fixed number of tool calls.

\subsubsection{Diagnostic assistance and verified recovery}

Read-only diagnosis is a useful operational boundary in its own right. An agent may localize a failure, identify a supported cause, or specify a discriminating test while an administrator retains responsibility for intervention. Diagnosis, proposed change, authorized execution, and verified recovery should therefore be recorded separately. Where changes are permitted, post-action observations should establish whether the affected service meets an explicit acceptance criterion; command completion alone does not demonstrate recovery. Evaluation should distinguish expert-confirmed diagnostic correctness, unsupported attribution, time to a useful finding, and operator effort from autonomous resolution. Monitoring should likewise distinguish process availability from successful evidence collection and delivery, so that an agent that continues running while its observation path fails is not reported as operationally healthy.

\subsection{KnowledgeOps for orchestrated autonomous pipelines}
\label{sec:knowledgeops-orchestrated-pipelines}

\subsubsection{Typed and scoped memory}

We use KnowledgeOps to describe the governed lifecycle of knowledge and memory consumed and produced by agents: capture, organization, retrieval, validation, reuse, revision, and retirement. \Gls{rag} is one access mechanism within this lifecycle, rather than the entire memory architecture. Useful memory distinguishes semantic notes, time-indexed episodes, derived state projections, versioned procedures, and agent journals. These represent different kinds of information: recorded knowledge or interpretations, what happened, the state inferred from recorded events, how a task should be performed, and a worker's account of its activity. Each should preserve its authorship, resource and project scope, evidence relationships, and applicable time or version. A source-grounded interpretation remains a derived claim until its support has been assessed; attaching a citation does not make the interpretation primary evidence.

Retrieval should follow the information need. Similarity search is useful for analogous incidents, identifier-based retrieval for known records, time- and entity-scoped queries for operational histories, and explicit aggregation over a defined population for counts and trends. Neither the number of relevant passages returned nor a sample of incidents establishes a facility-wide failure rate. Live operational state should be obtained from its authoritative service or a sufficiently recent observation, with any derived projection identified as such. Access checks should apply when memory is retrieved and reused; a shared summary must not bypass restrictions on its source records.

\subsubsection{Provenance, temporal validity, and concurrent updates}

Source identifiers and derivation relationships should survive summarization and agent handoffs. Several workers repeating an interpretation derived from one logbook entry do not provide independent corroboration. Whether a source exists, whether its revision is current, and whether it supports a particular claim are separate questions. When supporting records are corrected or superseded, dependent memories and procedures should be identified for revalidation or marked as requiring review, without erasing the historical record of their use.

Parallel workers may also observe different revisions or finish after an operator has corrected the shared state. Updates to a derived state should therefore identify the version on which they were based and reject or re-evaluate conflicting stale writes rather than accepting whichever response arrives last. Episodes should retain when an event occurred, when it was recorded, and the interval over which a claim applies. Derived projections should expose their freshness and the events through which they have been updated. These distinctions allow a later investigation to separate what the system knew when it acted from information recorded afterward; they do not make an incomplete observation history complete.

\subsubsection{From operational episodes to reusable procedures}

A completed workflow should retain an episode of what was attempted, the relevant conditions, its evidence, and the observed outcome, including failed or inconclusive approaches. This record may support a candidate procedure with explicit preconditions, required observations, expected outputs, and failure handling. Promotion into shared operational knowledge should follow the evaluation, expert review, versioning, and rollback process in Section~\ref{sec:tools-skills}, rather than rely on the agent's account of its own success. Appropriate checks may include deterministic analysis, fresh operational measurements, or expert acceptance. A procedure reused elsewhere must be checked for applicability to the receiving environment; a successful intervention on one host is not a universal remedy. This creates a controlled feedback path from execution to reusable knowledge without granting memory the authority to change production behavior.

\subsection{Facility-wide orchestration through a deterministic control plane}
\label{sec:unified-control-plane}

\subsubsection{A common interface over existing authorities}

A facility-wide control plane should provide a consistent way to discover capabilities, identify resources, submit bounded work, and inspect its outcome across otherwise disparate systems. This is a logical integration layer, not a requirement to replace existing control systems or concentrate all privileges in one service. Network management, Linux administration, storage, instrument control, computation, and knowledge services may retain their own execution engines and administrative boundaries. Reviewed deterministic adapters should translate common requests into the appropriate native interfaces, preserving each system's authorization, concurrency, and operational limits. A catalog or shared \gls{mcp} endpoint enables discovery and invocation but does not, by itself, establish common resource identity, action authority, or completion semantics.

The common contract should identify the target resource unambiguously, the requested capability and interface version, the authenticated caller, and the applicable scope. Resource relationships should support navigation across dependencies without assuming that names or identifiers from different systems are interchangeable. Unsupported capabilities and unresolved identities should be explicit failures, not occasions for the agent to guess a target or invent an alternative access path. The resulting control plane can be federated across independently deployed services while presenting a coherent interface to agents and operators. Table~\ref{tab:control-plane-responsibilities} summarizes the responsibility boundaries.

\begin{table}[t]
\centering
\caption{Recommended responsibility boundaries in a unified facility control plane. A common interface does not transfer all authority to the agent, orchestrator, or knowledge service.}
\label{tab:control-plane-responsibilities}
\begin{tabular}{p{0.22\textwidth}p{0.34\textwidth}p{0.34\textwidth}}
\hline
\textbf{Component} & \textbf{Primary responsibility} & \textbf{Boundary to preserve} \\
\hline
Agent worker & Interpret the task, investigate, and propose a plan or finding & Its narrative does not authorize an action or certify completion. \\
Orchestrator runtime & Maintain task identity, dependencies, budgets, handoffs, and acceptance state & Delegation does not enlarge the worker's permissions. \\
Deterministic adapters and execution services & Validate and dispatch requests; enforce local authority; report execution state & Native controllers and safety systems remain authoritative for effects. \\
KnowledgeOps service & Preserve scoped evidence references, episodes, interpretations, and reviewed procedures & Memory does not replace live state, execution records, or current approval. \\
\hline
\end{tabular}
\end{table}

\subsubsection{Executable contracts rather than prompt-only integration}

Repeated operations should be expressed as tested, versioned routines with validated parameters and explicit preconditions, following Section~\ref{sec:instrument-security}. The agent selects a capability and supplies bounded arguments; deterministic code resolves targets, validates the request, performs the approved sequence, and reports the result. Where a workflow is already established, its routine transitions should not require the model to reconstruct low-level commands on every run. Agent-generated integration code may be proposed and tested, but should enter the same reviewed release process as other production code rather than become a transient privileged execution path.

The contract should distinguish observation, change proposal, execution, status retrieval, and cancellation where the underlying service supports them. Approvals should be associated with the specific target, proposed change, and applicable validity conditions, and checked again before execution. Human confirmation may be presented centrally by the harness, as described above, while execution services enforce authenticated permissions and the applicable approval conditions. A script, notebook, or second agent must not bypass those conditions by using another path to the same resource. Returned observations should carry their source, time, and coverage; state-changing results should identify the accepted operation, execution state, affected resources, and available post-action evidence. Deterministic execution makes these rules explicit and testable; it does not guarantee that a distributed operation will succeed or that the agent selected the correct operation.

\subsubsection{Durable work and bounded orchestration}

An orchestrator may use agents to decompose goals, select procedures, or assess findings, but its runtime should retain durable task identity, dependency state, execution references, and acceptance conditions outside any worker's conversation. A handoff should carry the goal, resource scope, constraints, applicable procedure versions, resolvable evidence references, completed observations, and unresolved questions. Workers should receive only the capabilities and context required by their assignments. Replacing a worker or inference endpoint can then preserve the recorded basis of the work without implying equivalent model behavior or recovery of information that was never retained.

Parallel work should be bounded by service capacity, resource ownership, rate limits, and explicit time or cost budgets. Read-only investigations may proceed concurrently where supported, while conflicting changes require arbitration at the shared execution boundary. After interruption or an ambiguous response, the runtime should reconcile with the execution service before resubmission. Stable operation identifiers and idempotent bookkeeping do not guarantee exactly-once external effects; an operation spanning several systems may complete only partially. Such workflows need explicit partial-failure states and reviewed recovery or compensating actions where those actions are supported. A cancellation request should not be reported as completed cancellation without confirmation.

Acceptance should depend on task-specific evidence rather than only a worker's completion message. Deterministic checks, fresh service-level observations, and expert review should be applied according to the task, with bounded retry or escalation when acceptance criteria are not met. KnowledgeOps may retain the accepted outcome and propose a procedural update, but remembering a past approval must not authorize a later action. The control plane, orchestrator, and memory service thus support complementary parts of a continuous operational pipeline without conflating evidence, execution, and authority.

\subsubsection{Illustrative cross-domain investigation}

Consider several acquisition hosts reporting stalled transfers to a shared storage service. An orchestrator could create one incident with coordinated host, network-path, and storage investigations. Deterministic adapters would resolve resource identities and collect observations over compatible intervals. KnowledgeOps could retrieve a previous incident and its reviewed procedure as a source of hypotheses, not as the diagnosis of the current failure. If current evidence supports a storage-side access change as the cause, the diagnostic result could be accepted and a scoped change proposed to the responsible operator. The execution service would record the authorized intervention, while an independent service-level test would determine whether transfers recovered. The incident, unsuccessful hypotheses, source references, and verified outcome could then enter the governed knowledge lifecycle. This example illustrates the proposed separation of responsibilities, rather than reporting a measured autonomous recovery.

\section{Demonstration}
\label{sec:demonstration}

\subsection{\glsentryshort{eaa} at the 26-ID hard X-ray nanoprobe}

To anchor the recommendations above in concrete practice, we describe a representative deployment of \gls{eaa} at the 26-ID hard X-ray nanoprobe.

\subsubsection{System layout}
\label{sec:demo-architecture}

\begin{figure}[t]
    \centering
    \includegraphics[width=\textwidth]{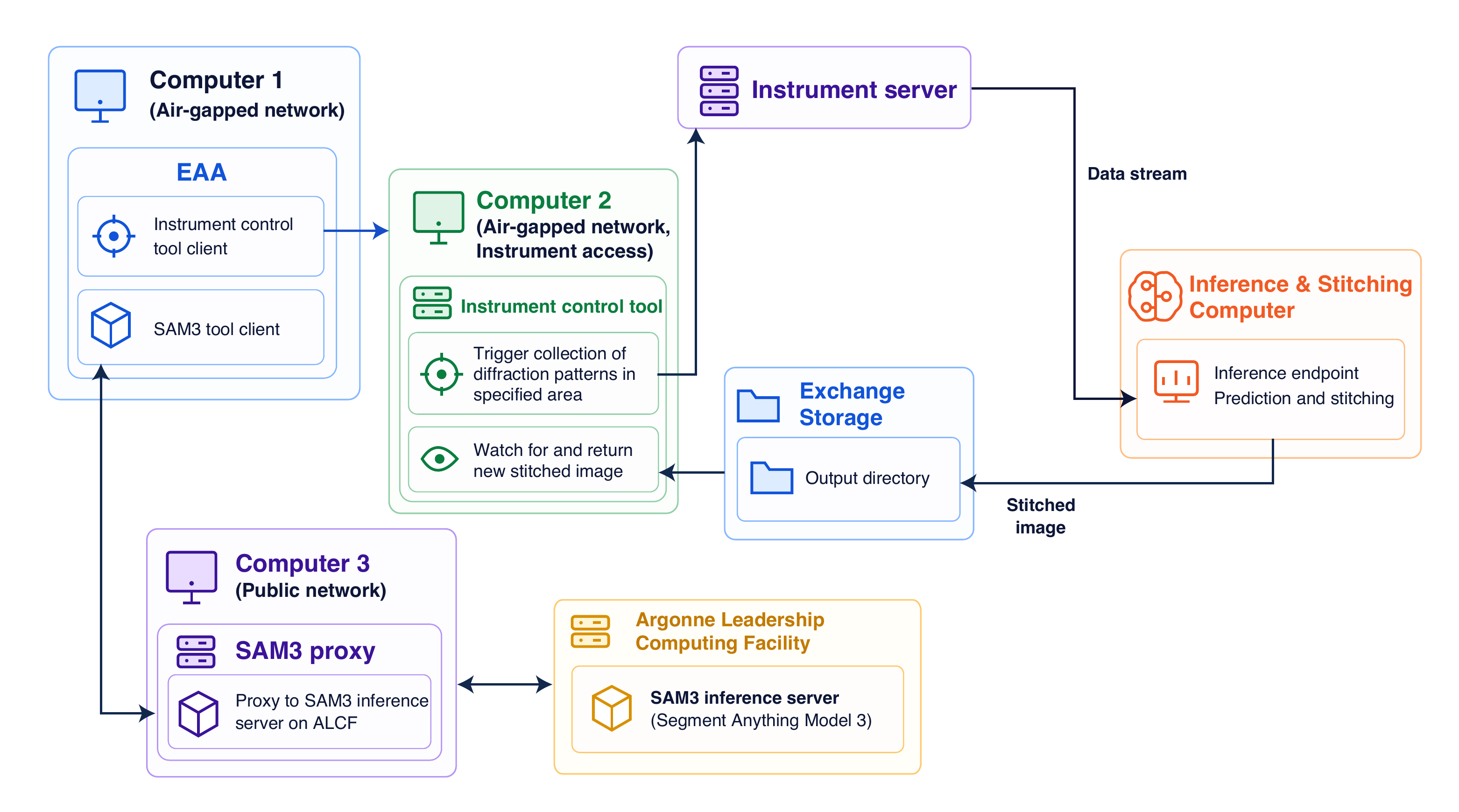}
    \caption{\Glsentryshort{eaa} deployment at the 26-ID hard X-ray nanoprobe. The beamline computer hosts \glsentryshort{eaa} along with an HTTP tool server; instrument control, online stitching, and heavy vision inference each run in their own processes or on their own computers, with shared NFS storage for data exchange. The pattern in Fig.~\ref{fig:tool-server-integration} is a generic schematic; this figure is one concrete instantiation of it.}
    \label{fig:s26-architecture}
\end{figure}

Fig.~\ref{fig:s26-architecture} shows the deployment. The agent harness (\gls{eaa}), the instrument-control tool server, and the \gls{sam3} proxy run on different computers because they have different environment and network-connectivity requirements. This follows the harness/tool-server decoupling principle in Section~\ref{sec:tool-servers-for-instrument-control}. The instrument-control tool server exposes the coarse/fine image acquisition tools (\texttt{acquire\_coarse\_scan} and \texttt{acquire\_fine\_scan}), receives tool calls from the agent, and executes data acquisition using the native control protocol for the beamline instruments. An inference and stitching computer hosts a deep learning-based ptychographic reconstruction model and image stitcher; the inference and stitching service listens to data streamed from the instrument server and is not directly controlled by the agent. These computers share the same NFS storage for data exchange, allowing the stitched ptychographic reconstruction image to be retrieved by the tool server and ultimately viewed by the agent. Image segmentation inference, \textit{e.g.}, \glsentrylong{sam3}, is offloaded to ALCF as described in Section~\ref{sec:compute-intensive}: a thin \glsentryshort{sam3} client on a public-network computer submits jobs to a \glsentryshort{sam3} server on ALCF. For light segmentation workloads, a fine-tuned \glsentryshort{sam3} model can perform inference on a local workstation. The 12-ID DAQ software captures live microscope camera feeds and transmits them for sample perimeter segmentation, confining the scanning range.

\subsubsection{An interactive session}
\label{sec:demo-session}

\begin{figure}[t]
    \centering
    \includegraphics[width=\textwidth]{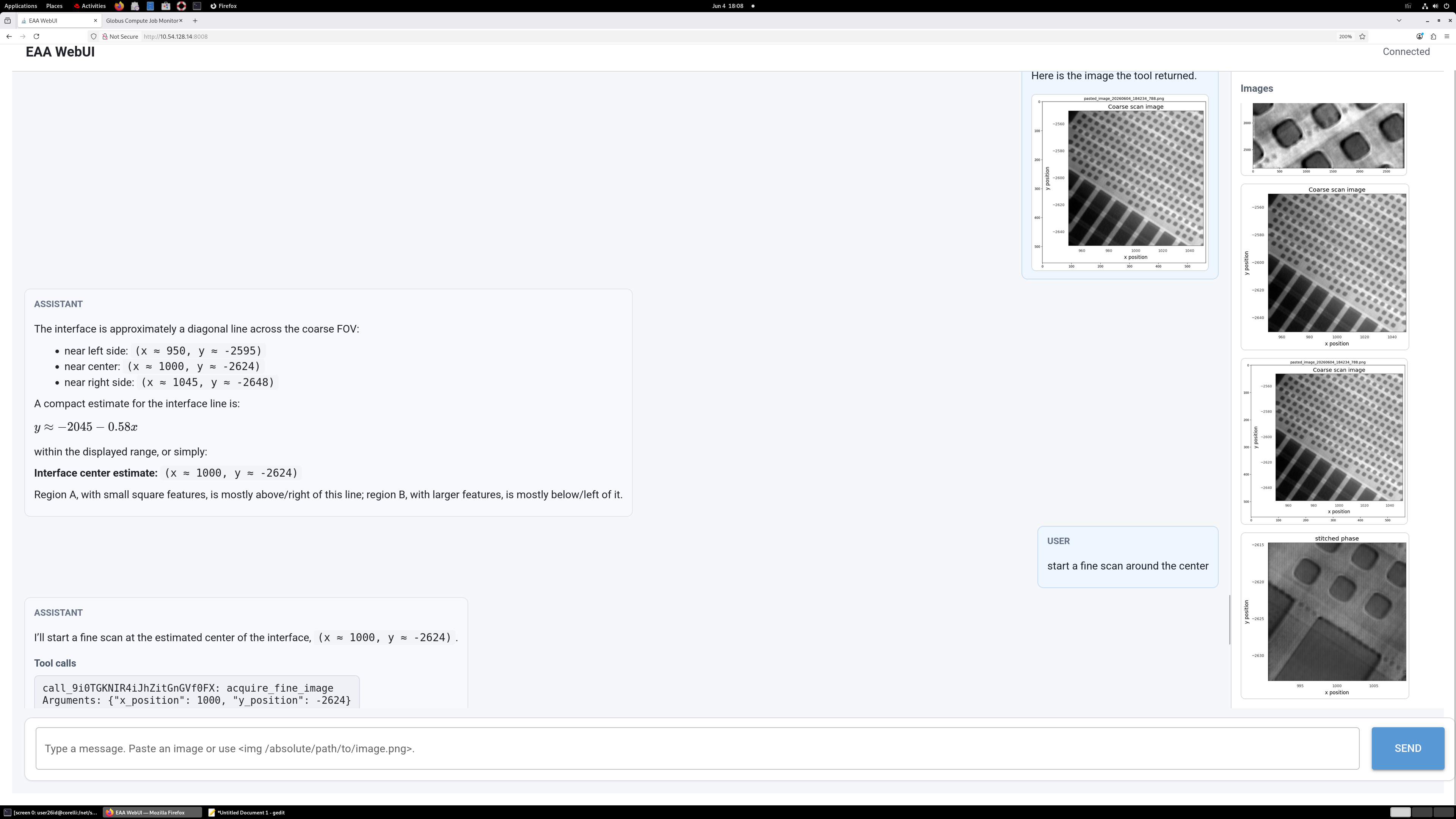}
    \caption{A representative interactive session with \glsentryshort{eaa} at the 26-ID nanoprobe. The user describes the desired measurement in natural language; the agent reasons over the overview images to localize an interface, computes the requested scan coordinates, and emits a single tool call to execute the line scan.}
    \label{fig:eaa-screenshot}
\end{figure}

Fig.~\ref{fig:eaa-screenshot} captures a representative user interaction with \gls{eaa} at 26-ID. The user is presented with previously acquired overview images of a patterned device that contains a roughly diagonal interface across the coarse field of view. The user asks the agent to ``start a fine scan around the center (of the interface).'' The agent inspects the overview images, identifies the interface, fits a line of the form $y = -2045 - 0.58\,x$ to it, computes the interface center at approximately $(x = 1000, y = -2624)$, and issues a single \texttt{acquire\_fine\_scan} tool call with those coordinates.

This single exchange exercises most of the recommendations in this paper. The overview images are tool-yielded multimodal data, handled per Section~\ref{sec:handling-tool-yielded-data}. The \texttt{acquire\_fine\_scan} call crosses the harness/tool-server boundary established in Section~\ref{sec:tool-servers-for-instrument-control}. On the foundation of these tools, an \texttt{roi-search-workflow} skill guides the agent to search for user-specified features in the sample. The skill serves as a high-level ``how-to'' document and suggests which tools to use, following the recommendation in Section~\ref{sec:tools-skills}. With appropriate combinations of tools, skills, and occasionally harness-specific workflows, more automated operations can be realized. For example, with a tool to move the zone plate, a skill documenting the procedure, and image-registration middleware in the harness, autonomous zone-plate focusing can be realized: the agent drives the focus sweep and selects the optimum from the resulting sharpness curve \cite{Du2026EAA}. The benefit extends beyond issuing a scan command more quickly. By connecting image evidence, calibration procedures, and instrument actions in a reviewable chain, the agent reduces the operator attention required for routine navigation and setup while preserving human authority over the scientific target and its interpretation.

\subsection{Osprey}
\label{sec:osprey}

The demonstration above is drawn from the beamline side of the facility. A parallel effort is under way on the accelerator side, where agents act on the \gls{aps} storage ring and injector rather than on an individual instrument. That effort is built on Osprey~\cite{Hellert2026Osprey}, an agentic platform for safety-critical control systems now deployed at several accelerators under the \gls{moat} seed project~\cite{Vay2026MOAT}. We summarize it here briefly because it was developed independently of the beamline work and reached a very similar deployment structure, suggesting that the principles in Section~\ref{sec:design-principles} are generally applicable to scientific instrument and facility control.

Osprey separates the same layers this paper recommends. The harness is currently a general coding agent driven through a web terminal, but the platform is harness- and model-agnostic by construction and has been tested both with frontier commercial models and with a locally hosted open-weight model, consistent with Section~\ref{sec:llm-endpoint}. Facility capabilities are exposed through \gls{mcp} servers rather than compiled into the harness, covering control-system access, archiver retrieval, workspace and artifact generation, and search over operational records.
Below the tool tier, a connector layer translates read, write, and discovery requests into the protocol of the local control system, so that \gls{epics} is one supported case rather than a built-in dependency. This is the same harness/tool-server decoupling described in Section~\ref{sec:tool-servers-for-instrument-control}, applied one level lower. Tool calls pass through a gate that performs command filtering, scope and limit checks, authorization, and log verification before reaching hardware, which places the guardrails of Section~\ref{sec:instrument-security} at the platform level rather than in each tool. Validated procedures are then captured as skills in the sense of Section~\ref{sec:tools-skills} and become invocable in a single sentence.
The reasoning behind this layering is that models and harnesses evolve at a timescale of weeks to months, whereas the work of connecting them to control systems, operating procedures, and facility knowledge should be stable and long-lasting. Osprey therefore treats that integration layer as the durable investment and allows the models above it to change independently.

The applications of Osprey at APS currently fall into three groups. The first is analysis of archived machine data.
One example is a survey of \gls{bpm}  \gls{ioc} timeouts, which drop the affected \glspl{bpm} out of orbit feedback. The agent worked back through the archiver to count how often each unit had failed over the past two years, separating transient glitches from genuine faults, and identifying the most problematic units for repair in the next maintenance cycle. Once the procedure had been checked by a system expert, it was packaged as a skill, and the same survey can be repeated on request.

The second group is on-demand diagnostics, where the relevant data does not exist until the agent asks for it. 
Power-supply current readbacks and \gls{bpm} turn-by-turn are sampled far too fast to archive continuously, so they are captured only on request.
One example is tracking down the source of \SI{60}{Hz} beam motion. The agent configured and fired the data-acquisition trigger, then searched the resulting waveforms for the \SI{60}{Hz} line component, finding both a pattern common across one magnet family and a single supply well above its peers that is now scheduled for inspection. The same workflow is being packaged as a skill, so it can be run for a specific investigation or on a routine monitoring schedule.
Another example is the commissioning study of the \gls{fofb} system. With feedback newly enabled in two sectors of the storage ring, the agent compared orbit motion before and after under similar beam conditions, and confirmed that there is clear improvement of beam orbit at those sectors, quickly accessing the performance of the newly commissioned system. 

The third group involves actuation. An early example is the beam emittance measurements at the photocathode gun and other locations in the injector. The agent drives an existing emittance measurement routine end to end, selecting the configuration, running the magnet scan, and returning fitted parameters with diagnostic plots. It does not generate the scan logic itself, which keeps the actuation deterministic and is the more robust arrangement wherever a validated routine exists. Osprey also allows the agent to compose control logic and write setpoints directly.

The accelerator does, however, lack the structural separation that organizes the beamline side of Fig.~\ref{fig:hub}. There is a single control system of a few million channels, and no natural per-instrument boundary along which to partition tool servers, skill sets, or approval policy. An agent acting on the machine acts on the one object that every beamline depends on.
The mitigation is gated write authority rather than isolation.
The large majority of the deployment is read-only, including archiver retrieval, channel discovery, and analysis. Where writes are permitted, it is checked against a whitelist of allowed channels and per-channel limits, with the write mode set per deployment.
At the \gls{aps}, actuation is currently confined to well-scoped measurement procedures such as the injector scan mentioned above.
Concurrency control is correspondingly looser than on the beamline side, where a Bluesky queue serializes access to an instrument. There is no equivalent single queueing authority for the machine, and the platform does not yet arbitrate between agents contending for the same subsystem.
With write access restricted as described this has not been a practical limitation, but explicit arbitration will be needed as autonomy and the number of concurrent sessions grow.

\section{Discussion and outlook}
\label{sec:outlook}

\paragraph{What has worked.} The single decision that has paid off most consistently is decoupling the agent harness from the instrument-control tools. We have swapped \gls{llm} endpoints, swapped harnesses, and modified skill sets repeatedly without touching the instrument-control code on the worker side. \Gls{mcp} and skills have provided a durable interoperability layer: tools written for one harness have been reusable in others with little or no modification. This architectural portability is necessary but not sufficient. Production systems should ultimately be evaluated by the quality and reproducibility of the scientific decisions they support, the time required to obtain interpretable evidence, the transfer of knowledge between users, and the operator time they free for scientific questions, rather than by throughput alone.

\paragraph{Evaluating operational and knowledge pipelines.} For the extensions in Sections~\ref{sec:aiops-facility-infrastructure}--\ref{sec:unified-control-plane}, evaluation should follow the chain from observation to accepted outcome. Relevant measures include evidence coverage, diagnostic correctness, time to a supported finding, operator effort, recovery verified at the affected service, and appropriate escalation. Knowledge reuse should be assessed for source support, temporal applicability, and preservation of access boundaries, including tests in which sources are corrected or workers resume with stale state. These are proposed evaluation dimensions, not measured results of a facility-wide autonomous deployment.

\paragraph{Immediate next steps.} The most foundational missing piece is the \gls{mcp}/Skills Hub itself. The central gateway in Fig.~\ref{fig:hub} is depicted as the spine of the architecture but does not yet exist. We aim to design and operate the hub in a lightweight, user-friendly way:
\begin{itemize}
    \item A GitHub organization will be created to host the source code and documentation for tool servers and skills. Each tool or skill will have its own repository.
    \item On top of that, we will develop a lightweight website resembling SkillsMP (\href{https://skillsmp.com/}{https://skillsmp.com/}) to serve as a catalog of tools and skills. The website will not hold the code or documents directly, but will link to the corresponding GitHub repositories. The website should:
    \begin{itemize}
        \item Allow users to browse or search for tools and skills by name or functional tag (\textit{e.g.}, \textit{microprobe}, \textit{database}, or \textit{ptychography})
        \item Provide installation and deployment instructions specific to each tool or skill
        \item Expose API endpoints and agent-friendly documentation that allow AI agents to search for, download, install, and deploy tools and skills.
    \end{itemize}
\end{itemize}

To protect the integrity of skill and \glsentryshort{mcp} repositories and prevent unauthorized changes, we will take the following measures:
\begin{itemize}
    \item The catalog website will be hosted inside the Argonne network and will not be publicly accessible.
    \item Each catalog entry will point to a specific commit or release tag rather than to a mutable repository branch. Updates will require an authenticated request through Argonne's single sign-on service, followed by administrator audit and approval. This process ensures that the reviewed version remains unchanged until a replacement is explicitly promoted.
\end{itemize}

The runtime control-plane extension should proceed separately from catalog publication. A staged integration can first expose read-only inventory, operational observations, and knowledge retrieval through common identity and result contracts, then add reviewable change proposals, and only subsequently admit authorized execution for selected procedures. Before a cross-domain workflow is enabled, each participating service should define its action scope, failure and cancellation behavior, arbitration boundary, and acceptance evidence. Shared memory should be introduced with the provenance and revision controls in Section~\ref{sec:knowledgeops-orchestrated-pipelines}. This sequence allows useful diagnostic assistance before broad write authority is available and does not require replacing the existing service backends.

Once the hub is operational, its four capability categories can be populated in parallel. Knowledge retrieval is the capability most readily available today. The APS retrieval-augmented-generation service (APS-\glsentryshort{rag}; \url{https://rag.aps.anl.gov/})~\cite{sainju2026apsrag} already searches machine logbooks, experiment and configuration records, work requests, operations discussions, technical reports, and other operational applications. The source of machine logbooks is \gls{bely}. Registering this service as an \glsentryshort{mcp} tool would make these sources available without requiring users to know the internal names of the underlying systems. However, knowledge exchange should not remain one-way. Corrections from scientists, annotated outcomes, and agent-proposed procedures should enter a governed write-back queue that records the source evidence, conditions of applicability, and review status before the information is promoted to durable facility knowledge.

Instrument control requires a shared skill registry so that a procedure validated at one beamline can be discovered, reviewed, and reused at another without being developed again. For instruments with an existing Bluesky deployment, \glsentryshort{bits} provides a uniform device registry and Queue Server interface, while \glsentryshort{bait-mcp} provides the corresponding tool-server entry point.

A first instance of such a registry is already starting to take shape on the accelerator side. The effort described in Section~\ref{sec:osprey} has established a marketplace of shared accelerator skills, populated by workflows that were validated in operation and then packaged for one-sentence reuse.
It serves the same function as the Skills Hub in Fig.~\ref{fig:hub}.

The analysis and computing capability is also partially in place because \gls{eaa} offloads \gls{sam3} inference and Pty-Chi reconstruction to Polaris at the \glsentryshort{alcf}. These services should be made available through the hub so that other harnesses can access them through a common interface. In addition, a cross-cutting observability service should record tool calls, reasoning traces, evidence, and confirmed human overrides in a common, queryable format. The Hutch package~\cite{yin_hutch} is a first step in this direction.

We intend the beamline and accelerator catalogs to cross-reference one another. Treating these systems as a unified framework enables complex, automated workflows where beamline users gain selective local control over specific accelerator parameters. For instance, users could implement local orbit bumps directly or query real-time accelerator diagnostics to correlate beam performance with X-ray-based experimental measurements.
Their contents generalize in different directions. A beamline skill often transfers to another beamline with a similar experimental setup, and increasingly to a comparable instrument at another facility, so the natural scope for that catalog is the technique rather than the site. What the two do share is a set of capability classes, including channel discovery, archiver retrieval, logbook and document search, plot generation, and the observability and provenance services described above. 
Cross-referencing matters for questions that span both domains, such as beam stability at a source point or drift correlated with machine state, where an agent needs to discover a machine-side diagnostic skill and invoke it read-only without inheriting write access to the accelerator.

\paragraph{Longer-term outlook.}
The longer-term opportunity is to increase the scientific value of instruments throughout their operational lifetimes. Agents can perform routine calibration, measurement execution, quality monitoring, and initial data synthesis, while scientists retain authority over experimental goals, uncertainty, anomalies, and interpretation. Knowledge can then flow in both directions: scientists teach procedures and judgment to the system, while the system supports future users by returning explanations, provenance, and lessons accumulated across experiments. Validated lessons can become versioned skills that move between instruments and, eventually, between facilities; unvalidated experiences should remain isolated as candidate knowledge. Specific implementations will change, but separability, safety, provenance, human authority, and governed learning will remain durable requirements. Success should therefore be measured not only by throughput, but also by reproducibility, access, the time required to obtain trustworthy evidence, and the quality of the scientific decisions that the system enables.

\section*{Acknowledgements}

This work was supported by the Advanced Photon Source, a U.S.~Department of Energy (DOE) Office of Science user facility at Argonne National Laboratory, and is based on research supported by the U.S.~DOE Office of Science-Basic Energy Sciences, under Contract No.~DE-AC02-06CH11357.

\bibliographystyle{unsrt}
\bibliography{refs}

\newpage

\renewcommand{\thefigure}{A\arabic{figure}}
\setcounter{figure}{0}

\section*{Appendix A. Screenshots of user interfaces of recommended agentic harnesses}

\begin{figure}[H]
    \centering
    \includegraphics[width=1\linewidth]{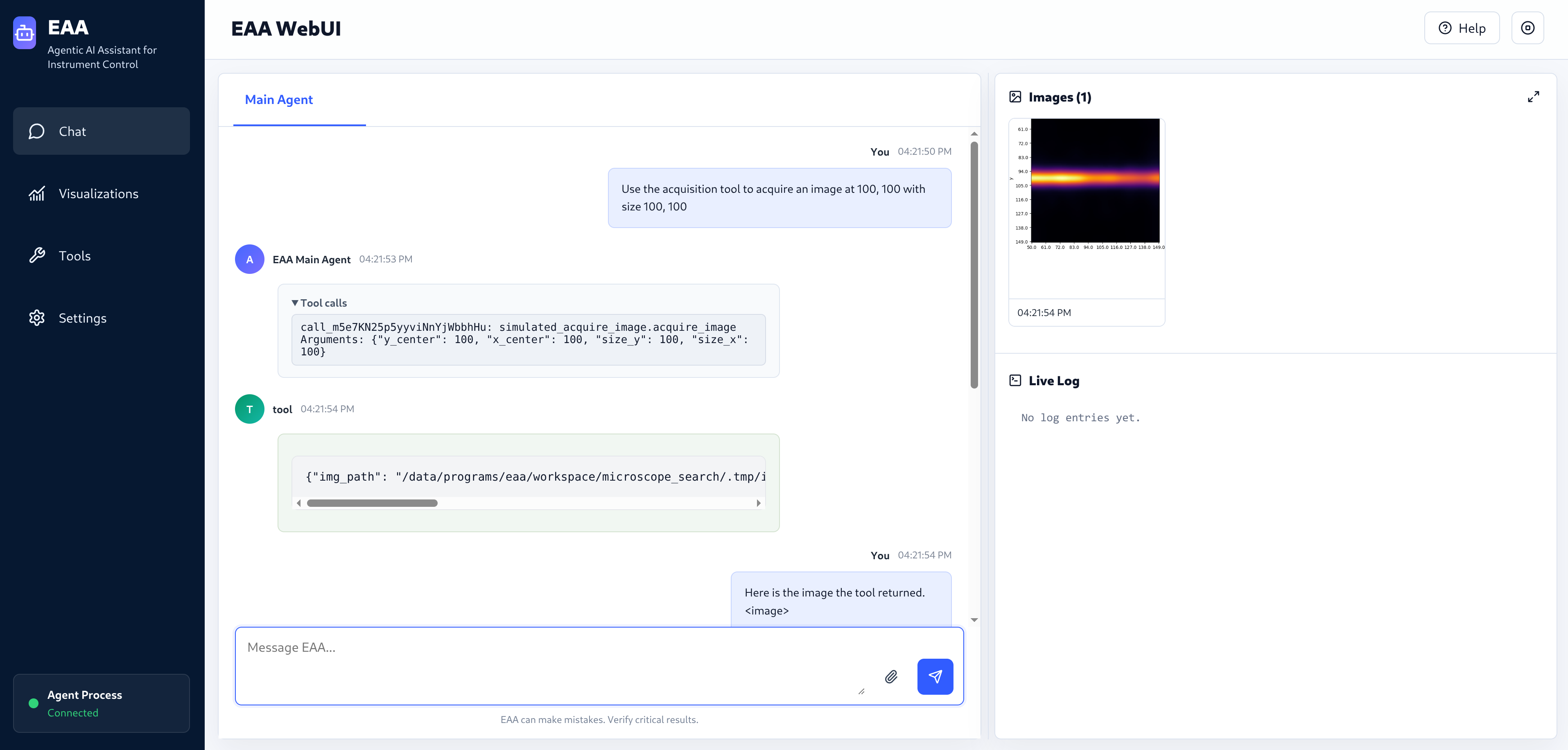}
    \caption{Screenshot of \glsentryshort{eaa}'s WebUI.}
    \label{fig:screenshot_eaa}
\end{figure}

\begin{figure}[H]
    \centering
    \includegraphics[width=1\linewidth]{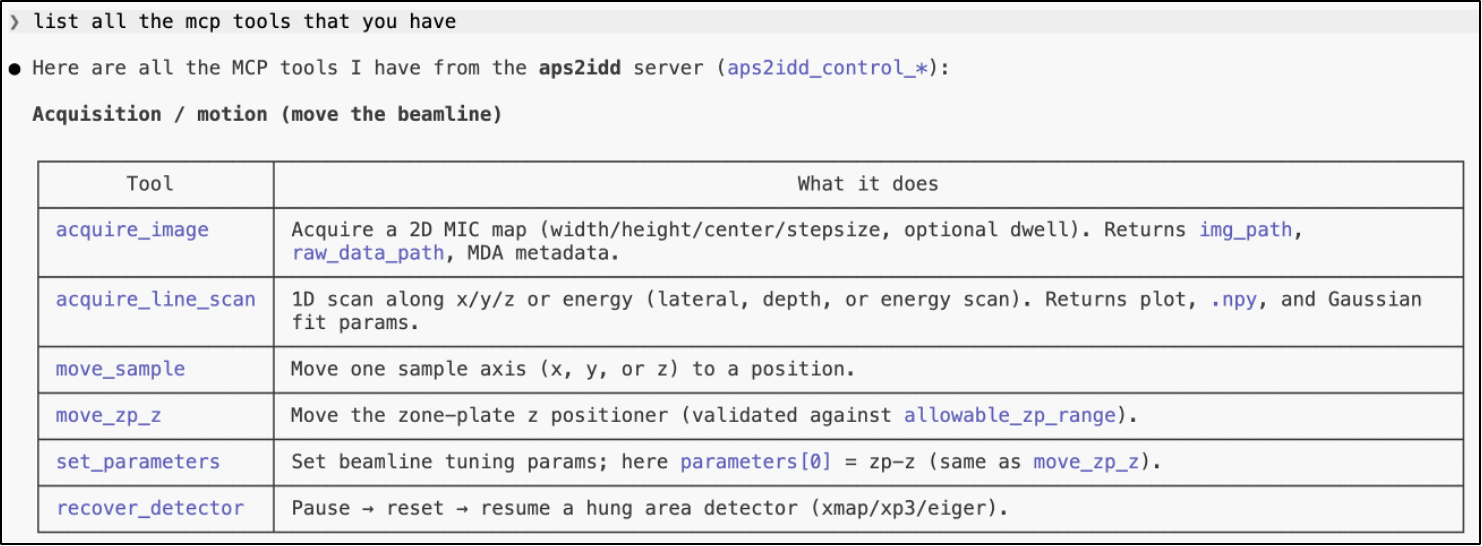}
    \caption{List of \glsentryshort{mcp} tools that can be launched in Claude Code.}
    \label{fig:screenshot_cc_mcp_1}
\end{figure}

\begin{figure}[H]
    \centering
    \includegraphics[width=1\linewidth]{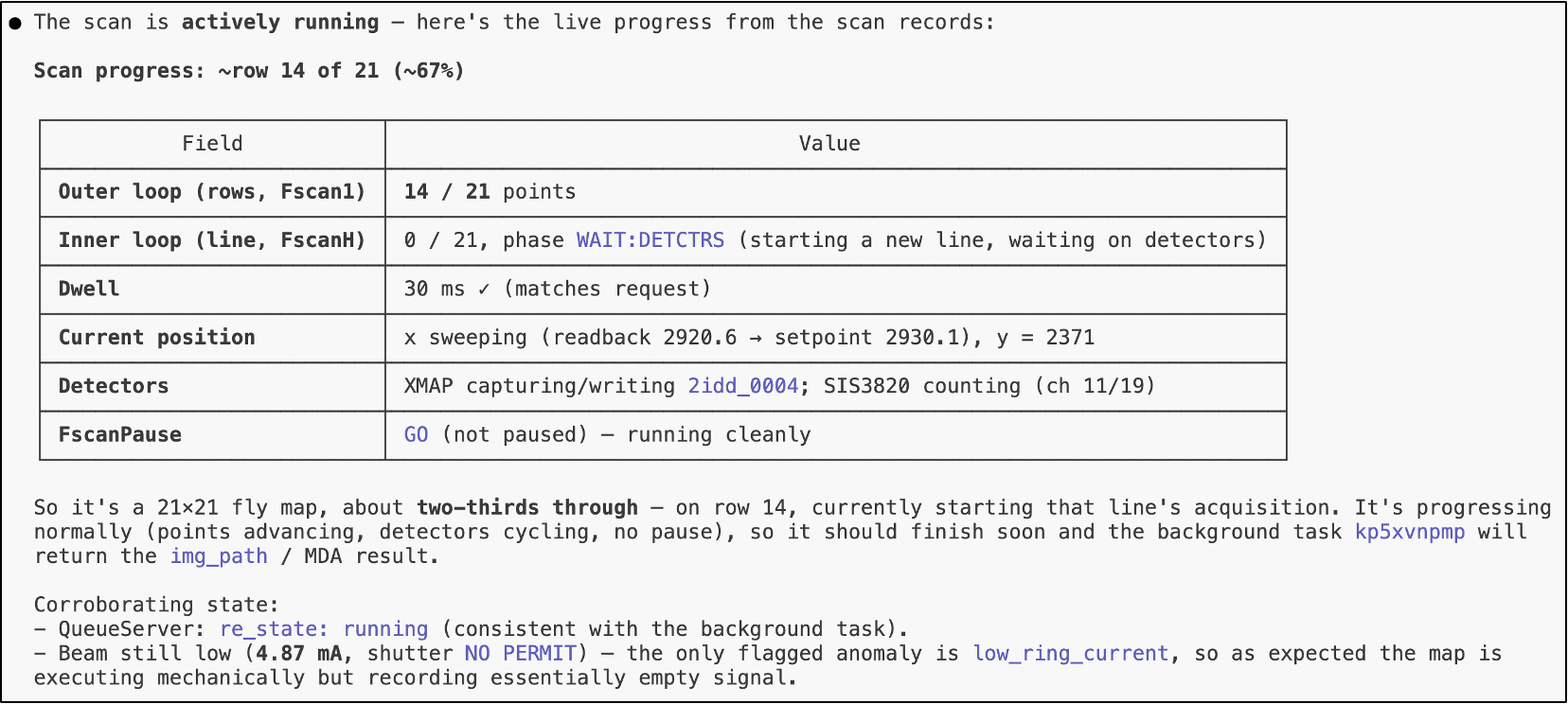}
    \caption{Requesting scan progress via \glsentryshort{mcp} tool call in Claude Code.}
    \label{fig:screenshot_cc_mcp_2}
\end{figure}

\begin{figure}[H]
    \centering
    \includegraphics[width=1\linewidth]{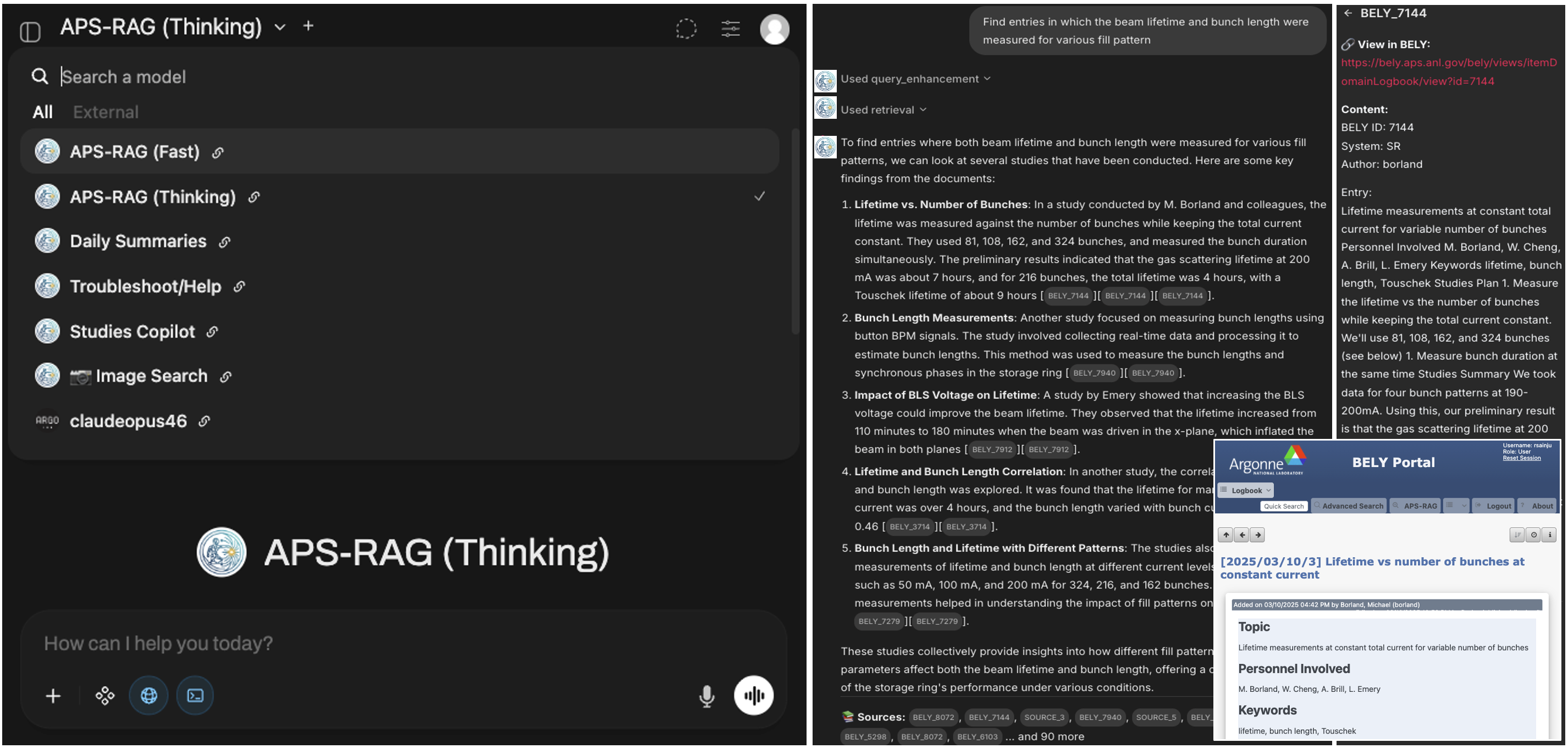}
    \caption{Web interface of the APS retrieval-augmented-generation service. The left panel shows the available modes and assistants, including daily summaries, troubleshooting, and image search. The right panel shows a query about beam-lifetime and bunch-length measurements. Selecting an inline citation opens the corresponding \gls{bely} entry in a sidebar, preserving the connection between the generated answer and its operational source.}
    \label{fig:screenshot_apsrag_ui}
\end{figure}

\begin{figure}[H]
    \centering
    \includegraphics[width=1\linewidth]{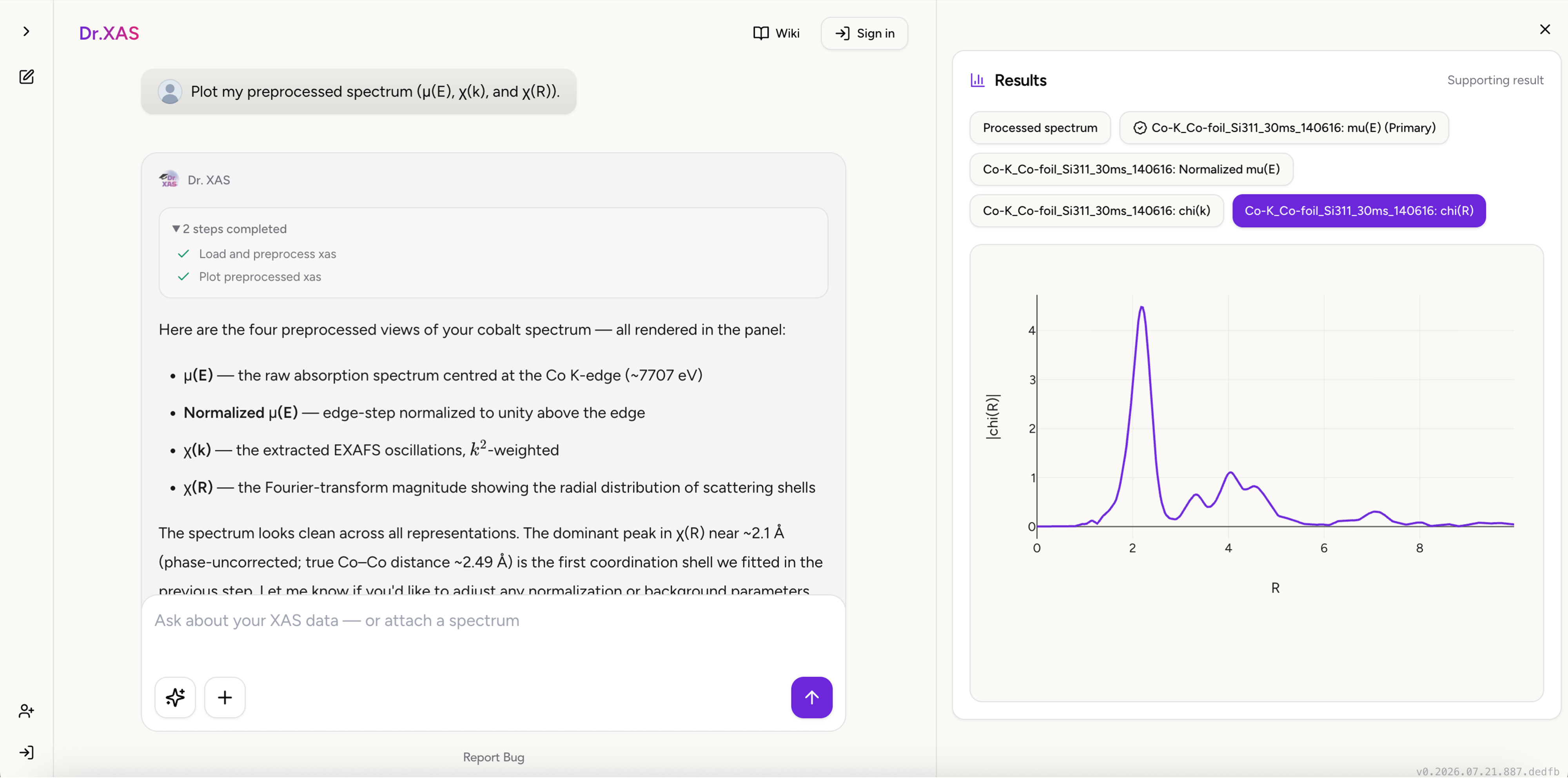}
    \caption{Screenshot of the Dr.~\glsentryshort{xas} WebUI for spectroscopy analysis.}
    \label{fig:drxas-screenshot}
\end{figure}

\begin{figure}[H]
    \centering
    \includegraphics[width=1\linewidth]{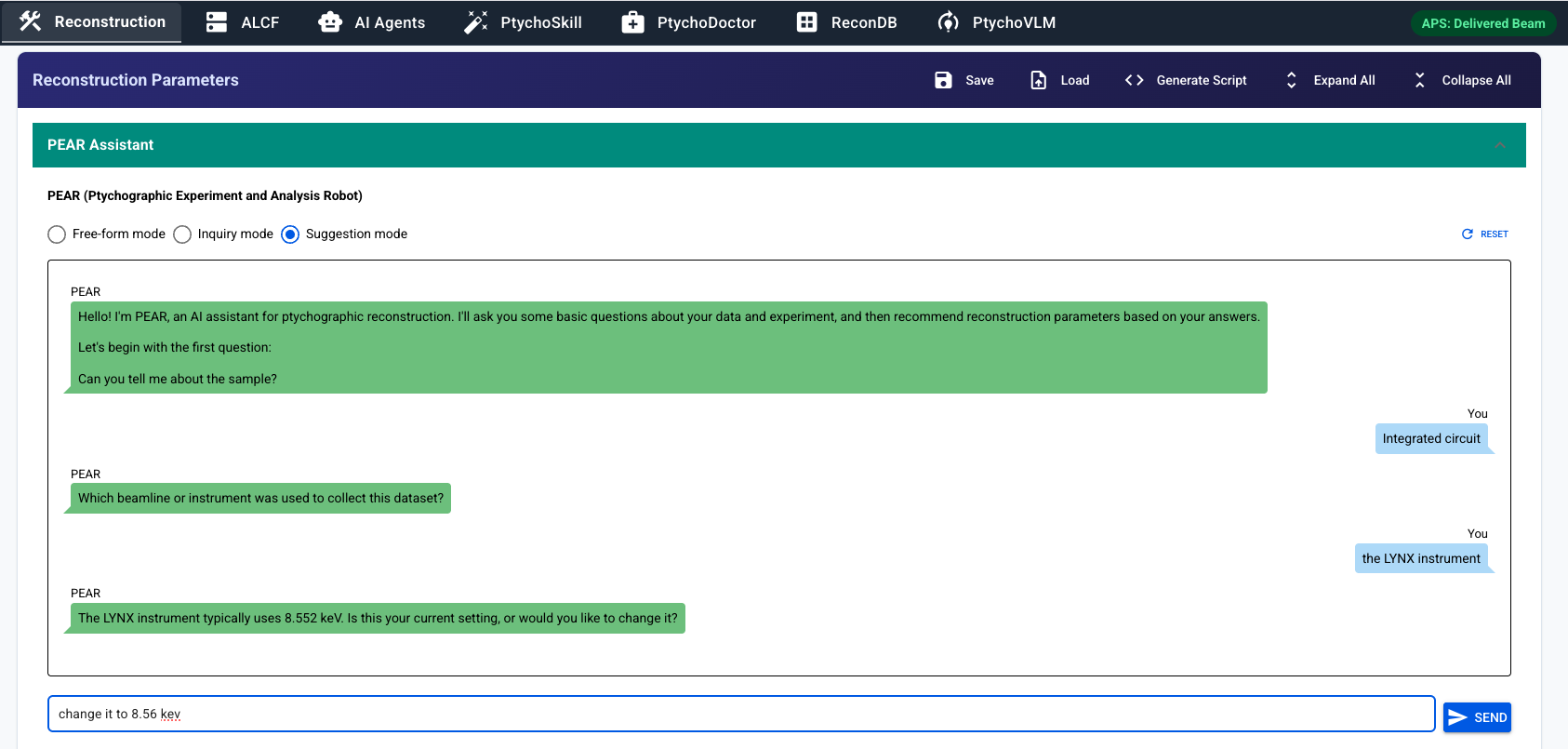}
    \caption{Screenshot of the PEAR WebUI for automated ptychography reconstruction.}
    \label{fig:pear-screenshot}
\end{figure}

\begin{figure}[H]
    \centering
    \includegraphics[width=1\linewidth]{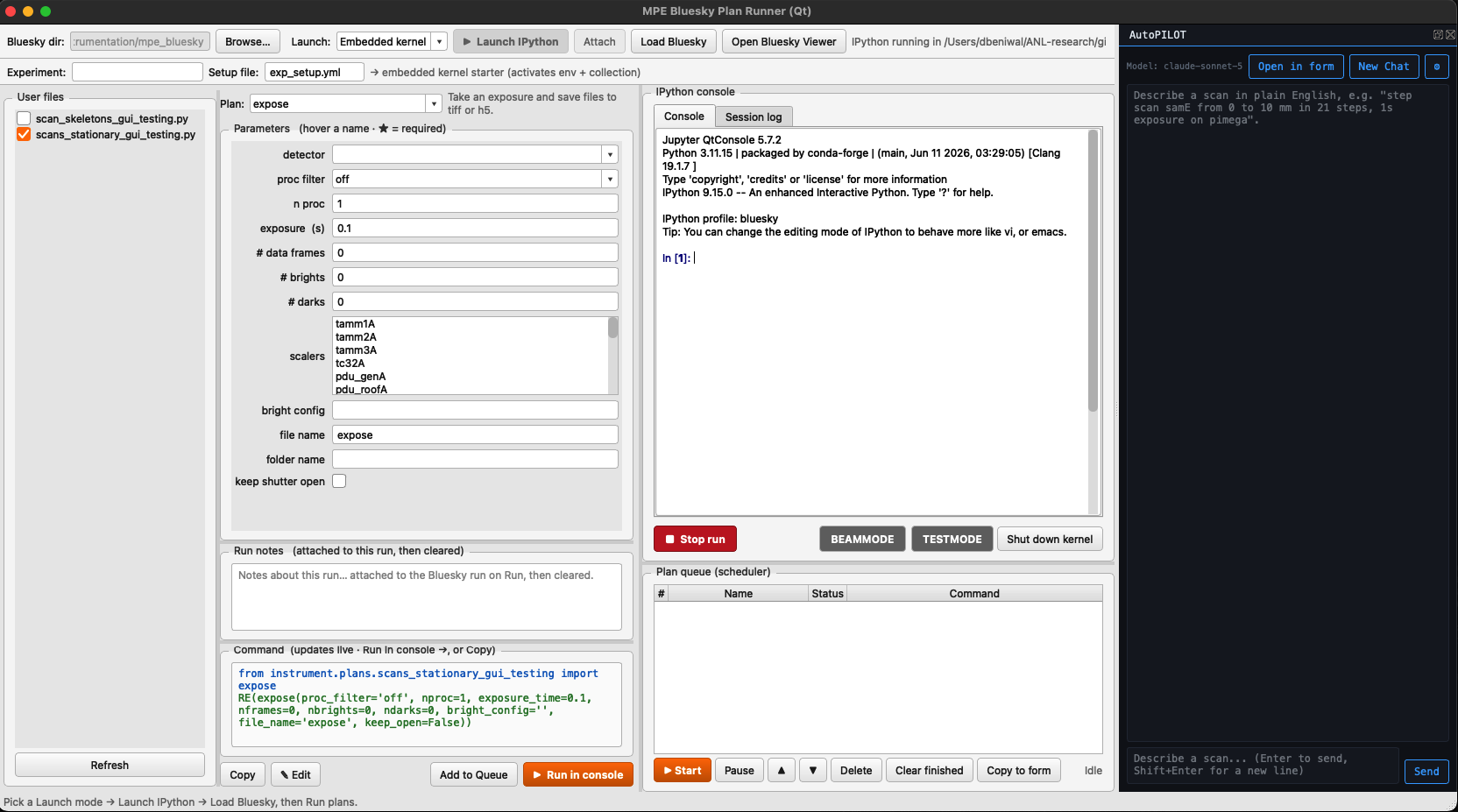}
    \caption{Screenshot of the Bluesky-Plan Interface for Launch, Operation and Tracking (B-PILOT) software with the embedded AutoPILOT panel that provides an agentic interface to B-PILOT.}
    \label{fig:screenshot_bpilot}
\end{figure}

\begin{figure}[H]
    \centering
    \includegraphics[width=1\linewidth]{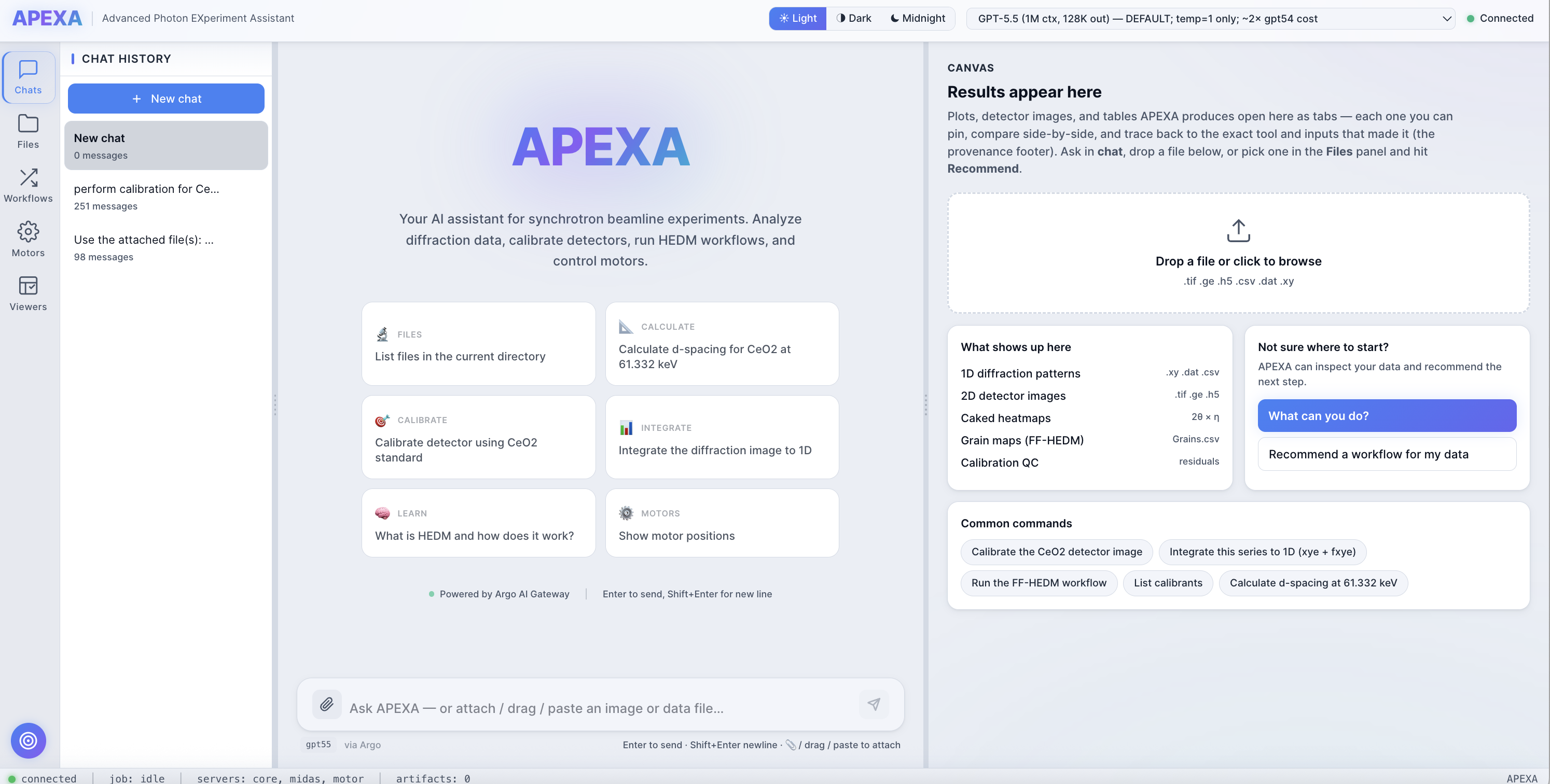}
    \caption{Web interface of APEXA, an Advanced Photon EXperiment Assistant for \glsentryshort{hedm} analysis.}
    \label{fig:APEXA-GUI}
\end{figure}

\begin{figure}[H]
    \centering
    \includegraphics[width=1\linewidth]{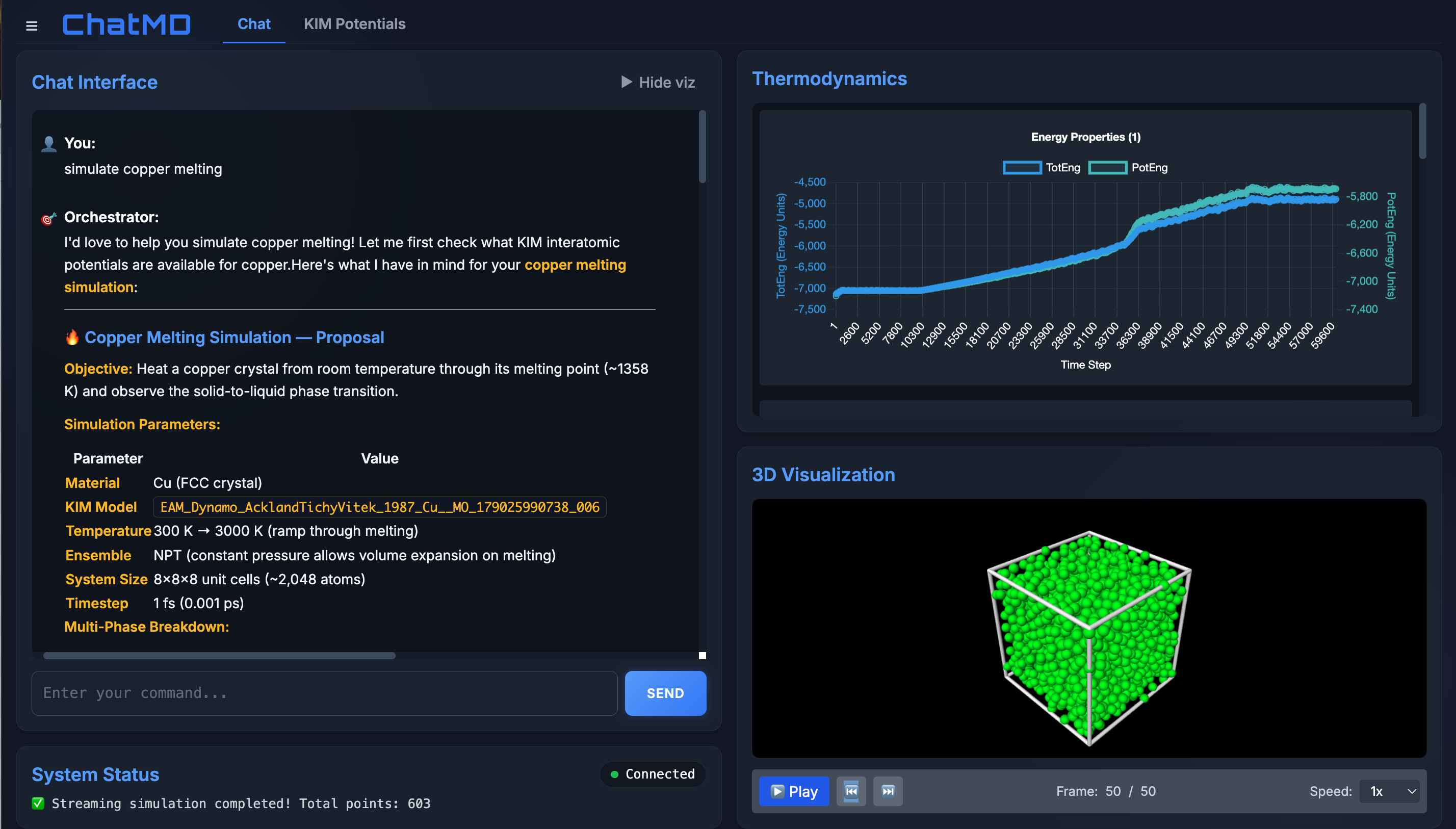}
    \caption{ChatMD-WebUI: An Agentic Workflow for Molecular Dynamics Simulations.}
    \label{fig:ChatMD_GUI}
\end{figure}

\begin{figure}[H]
    \centering
    \includegraphics[width=1\linewidth]{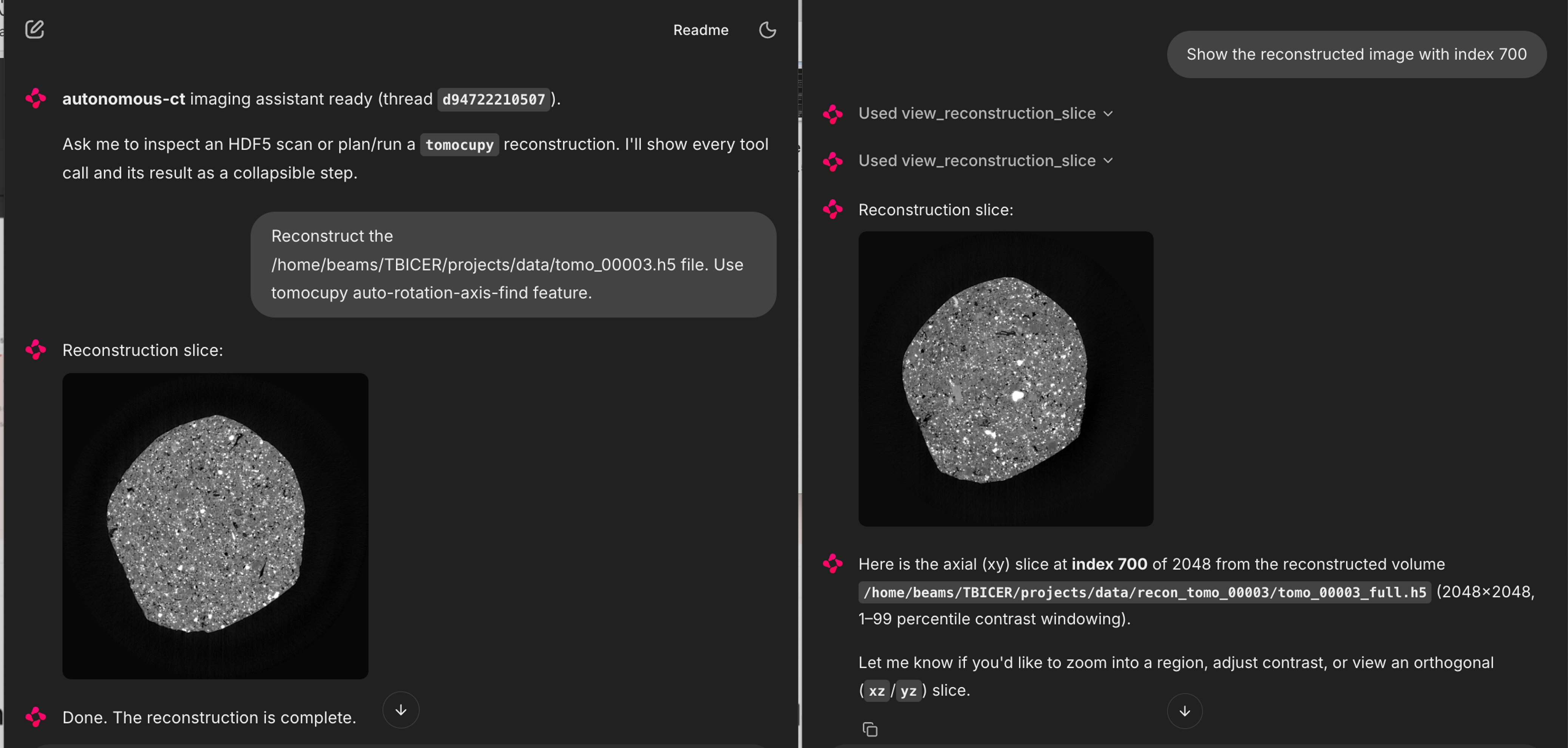}
    \caption{Auto-CT is an agentic system for autonomous analysis of tomography datasets. It currently uses tomocupy as a reconstruction engine, and supports visualization and manipulation of reconstructed slices.}
    \label{fig:autoct_GUI}
\end{figure}

\begin{figure}[H]
    \centering
    \includegraphics[width=1\linewidth]{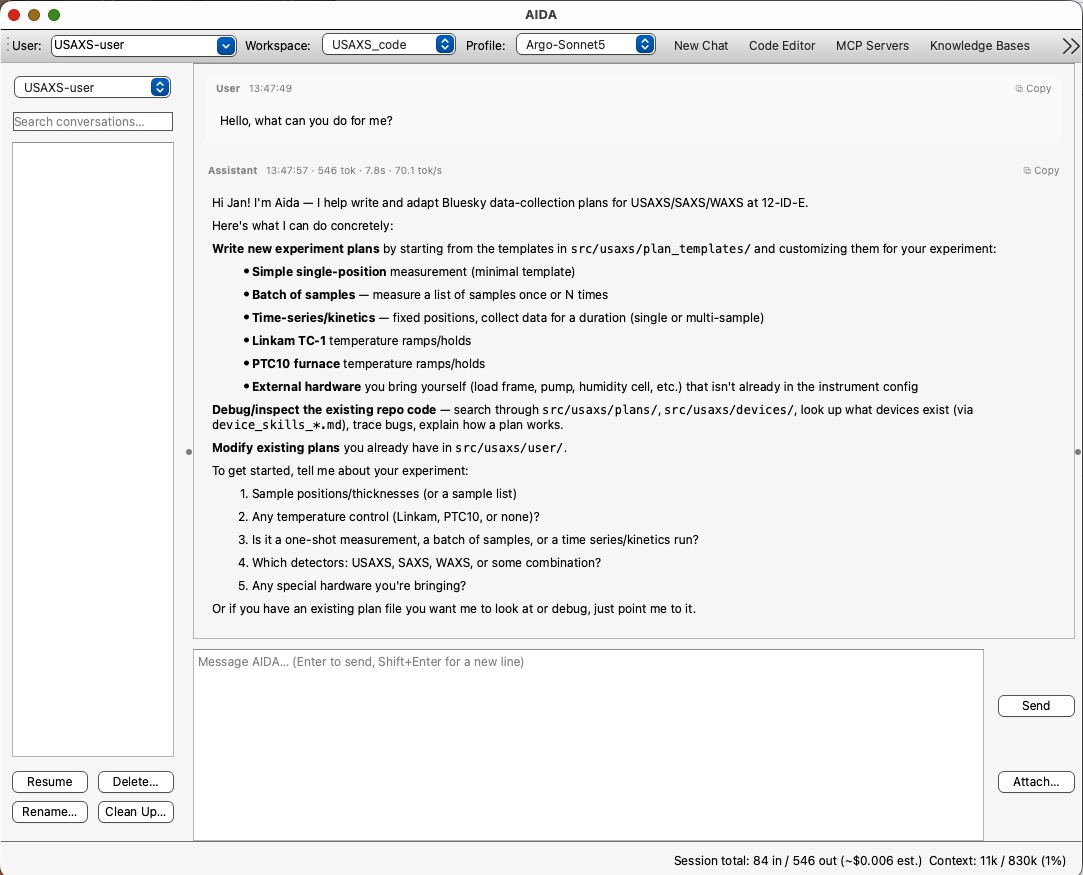}
    \caption{\Gls{aida} interface for Bluesky data-collection planning at the APS USAXS beamline.}
    \label{fig:aida_usaxs}
\end{figure}

\end{document}